\documentclass[prd,twocolumn,eqsecnum,showpacs,amsfonts,amssymb]{revtex4}

\usepackage{graphicx}

\usepackage{bm}

\newcommand{\beq}{\begin{equation}}
\newcommand{\eeq}{\end{equation}}
\newcommand{\beqs}{\begin{eqnarray}}
\newcommand{\eeqs}{\end{eqnarray}}
\newcommand{\lsim}{\mathrel{\raisebox{-
.6ex}{$\stackrel{\textstyle<}{\sim}$}}}
\newcommand{\gsim}{\mathrel{\raisebox{-
.6ex}{$\stackrel{\textstyle>}{\sim}$}}}

\begin{document}

\title{An Analysis of Scheme Transformations in the Vicinity of an Infrared 
Fixed Point} 

\author{Thomas A. Ryttov$^a$ and Robert Shrock$^b$}

\affiliation{(a) \ Jefferson Physical Laboratory \\ 
Physics Department, Harvard University,  \\
Cambridge, MA  02138}

\affiliation{(b) \ C. N. Yang Institute for Theoretical Physics \\
Stony Brook University \\
Stony Brook, NY 11794 }

\begin{abstract}

We give a detailed analysis of the effects of scheme transformations in the
vicinity of an exact or approximate infrared fixed point in an asymptotically
free gauge theory with fermions.  We list necessary conditions that such
transformations must obey and show that, although these can easily be satisfied
in the vicinity of an ultraviolet fixed point, they constitute significant
restrictions on scheme transformations at an infrared fixed point. We
construct acceptable scheme transformations and use these to study the
scheme-dependence of an infrared fixed point, making comparison with our
previous three-loop and four-loop calculations of the location of this point in
the $\overline{MS}$ scheme.  We also use an illustrative hypothetical exact
$\beta$ function to investigate how accurately analyses of finite-order series
expansions probe an infrared fixed point and the effect of a scheme
transformation on these.  Some implications of our work are discussed.

\end{abstract}


\maketitle


\section{Introduction}
\label{intro}

The evolution of an asymptotically free gauge theory from the weakly coupled
ultraviolet (UV) regime to the infrared (IR) regime is of fundamental interest.
Here we study this evolution for a theory with gauge group $G$ and a given
content of massless fermions.  The UV to IR evolution is determined by the
renormalization group $\beta$ function of the theory, which describes the
dependence of $g \equiv g(\mu)$, the running gauge coupling, on the Euclidean
momentum scale, $\mu$ \cite{rg}.  If a theory is asymptotically free, with a
small gauge coupling at a high scale $\mu$, and if the $\beta$ function of this
theory has a zero at a value $\alpha_{IR}$, then as the scale $\mu$ decreases
from large values, the coupling evolves toward $\alpha_{IR}$, which is thus an
exact or approximate infrared fixed point (IRFP) of the renormalization group
\cite{nonsingular}.  The approximate determination of the location of
$\alpha_{IR}$ from a perturbative calculation of $\beta$ is complicated by the
fact that at three-loop and higher order, the $\beta$ function is dependent
upon the scheme used for the regularization and renormalization of the theory.
It is clearly important to assess the effect of this scheme dependence on the
determination of $\alpha_{IR}$. This can be done by calculating $\beta$ in one
scheme, performing a transformation to another scheme, and comparing the
respective values of $\alpha_{IR}$ in these schemes.  In Ref. \cite{scc} we
pointed out that there is far less freedom in choosing scheme transformations
at an IR fixed point than there is at a UV fixed point (UVFP), and we reported
results from a study of scheme dependence in the calculation of an IR fixed
point to three-loop and four-loop order.  Since the one-loop and two-loop terms
in the $\beta$ function are scheme-independent, with scheme-dependence entering
only at the level of three loops and higher, one plausibly expects that if
$\alpha_{IR}$ is small, reasonably well-behaved scheme transformations should
not shift it very much, and our results confirm this expectation.  However,
these transformations do have a significant effect when $\alpha_{IR}$ is of
order unity, as is generically the case when one is investigating the boundary,
as a function of the number of fermions, between the infrared phases with and
without spontaneous chiral symmetry breaking.

In this paper we present a detailed analysis of scheme transformations in the
vicinity of an IR fixed point. We focus mainly on vectorial gauge theories but
also remark on chiral gauge theories. For a vectorial gauge theory, it
is straightforward to generalize our assumption of massless fermions to the
case of finite-mass fermions; essentially, by use of the decoupling theorem
\cite{ac}, at a given scale $\mu$, one includes the subset of the fermions with
masses small compared with $\mu$ and integrates out those with masses greater
than $\mu$.  In contrast, for a chiral gauge theory, the gauge invariance
requires massless fermions.  As an input to our present work, we use our
previous calculations of IR zeros of $\beta$ to three-loop and four-loop order
in the $\overline{MS}$ scheme \cite{bvh} (see also \cite{ps}, the
results of which agree with \cite{bvh}).  

We define $\alpha=g^2/(4\pi)$, $a \equiv g^2/(16\pi^2) = \alpha/(4\pi)$, and
\beq
\beta_\alpha \equiv \frac{d\alpha}{dt} \ , 
\label{betadef}
\eeq
where $t=\ln \mu$. This has the series expansion
\beq
\beta_\alpha = -2\alpha \sum_{\ell=1}^\infty b_\ell \, a^\ell = 
 -2\alpha \sum_{\ell=1}^\infty \bar b_\ell \, \alpha^\ell \ , 
\label{beta}
\eeq
where $\bar b_\ell = b_\ell/(4\pi)^\ell$.  The coefficients $b_1$ and $b_2$
were calculated in \cite{b1} and \cite{b2}.  The $b_\ell$ for $\ell=1,2$ are
independent of the scheme used for regularization and renormalization, while
$b_\ell$ with $\ell \ge 3$ are scheme-dependent \cite{gross75}.  One scheme
involves dimensional regularization \cite{dimreg} and minimal subtraction
($MS$) of the poles at dimension $d=4$ in the resultant Euler $\Gamma$
functions \cite{ms}.  The heavily used modified minimal subtraction
($\overline{MS}$) scheme also subtracts certain related constants
\cite{msbar}. Calculations of $b_3$ and $b_4$ in the $\overline{MS}$ scheme
were given in \cite{b3,b4}.  Just as the calculation of $b_1$ and demonstration
that $b_1 > 0$ was pivotal for the approximate Bjorken scaling observed in deep
inelastic electron scattering at SLAC and the development of quantum
chromodynamics (QCD) \cite{b1,b1b2}, the computation of $b_\ell$ for
$\ell=2,3,4$ has been important for many QCD calculations and fits to
experimental data, including data on $\alpha_s(Q)$ \cite{bethke}.  Thus,
although the expansion (\ref{beta}) is not a Taylor-series expansion with a
finite radius of convergence, but instead is only an asymptotic series
\cite{dyson} and neglects nonperturbative effects such as instantons
\cite{cdg}, comparisons of finite-order calculations with
experimental data in QCD at momentum scales large compared with the confinement
scale, $\Lambda_{QCD} \simeq 300$ MeV, have shown that one can reliably use the
perturbative $\beta$ function in the deep Euclidean regime.

In the vicinity of the UV fixed point at $\alpha=0$, one can carry out a scheme
transformation that renders three- and higher-loop terms zero \cite{thooft77}.
Below we will present an explicit construction of a scheme transformation that
achieves this.  Considerable work has been done on scheme (and related scale)
transformations that reduce higher-order corrections in QCD calculations
\cite{stevenson}-\cite{kataev}. However, as we showed in \cite{scc}, in order
to be acceptable, a scheme transformation must satisfy a number of conditions,
and although these can easily be satisfied in the vicinity of a UV fixed point,
they are highly nontrivial, and are strong restrictions, in the vicinity of an
IR fixed point.  This is especially true when $\alpha_{IR}$ grows to a value 
of order unity, so that infrared theory is becoming strongly coupled.


\section{Background}
\label{background}

We first recall some relevant background.  As noted above, except where
otherwise indicated, we will consider a vectorial non-Abelian gauge theory.
This theory has gauge group $G$ and $N_f$ massless (Dirac) fermions
transforming according to a representation $R$.  For a given $G$ and $R$, as
$N_f$ increases, $b_1$ decreases and eventually vanishes at
\cite{casimir,nfreal}
\beq
N_{f,b1z} = \frac{11C_A}{4T_f} \ .
\label{nfb1z}
\eeq
Since we restrict our considerations to an asymptotically free theory, we
require that, with our sign conventions, $b_1 > 0$, which implies an upper
limit on $N_f$, namely, 
\beq
N_f < N_{f,max} \equiv N_{f,b1z} \ . 
\label{nfrange}
\eeq
If $N_f$ is zero or sufficiently small, then $b_2$ has the same positive sign
as $b_1$, so $\beta$ has no (perturbative) IR zero for $\alpha \ne 0$
\cite{nonpertzero}.  With a sufficient increase in $N_f$, $b_2$ vanishes. 
This occurs at
\beq
N_{f,b2z} = \frac{17 C_A^2}{2T_f(5C_A+3C_f)} \ .
\label{nfb2z}
\eeq
For $N_f > N_{f,b2z}$, $b_2$ reverses sign, becoming negative.  Since
$N_{f,b2z} < N_{f,max}$, it follows that in the interval $I$ defined by
\beq
I: \quad N_{f,b2z} < N_f < N_{f,max} \ , 
\label{nfinterval}
\eeq
the two-loop $\beta$ function has an IR zero at $a_{IR,2\ell}=-b_1/b_2$,
i.e. 
\beq
\alpha_{IR,2\ell} = - \frac{4\pi b_1}{b_2} \ , 
\label{alfir_2loop}
\eeq
which is physical for $b_2 < 0$. Since $b_1$ and $b_2$ are scheme-independent,
so is $\alpha_{IR,2\ell}$. In contrast, an IR zero calculated at $n$-loop
($\ell$) order for $n \ge 3$ is dependent upon the scheme $S$ used for the
calculation, so we denote it here as $\alpha_{IR,n\ell,S}$.  (In \cite{bvh} we
denoted this simply as $\alpha_{IR,n\ell}$ since we were working there entirely
in the context of the $\overline{MS}$ scheme.) For a given gauge group $G$ and
fermion representation $R$ (provided that $N_f \in I$, so that the two-loop
$\beta$ function has a zero),
\beq
\alpha_{IR,2\ell} \  {\rm is \ a \ decreasing \ function \ of} \  N_f.
\label{alfir_2loop_nfdependence}
\eeq
As $N_f$ approaches $N_{f,max}$ from below, $b_1 \to 0^+$, while $b_2$
approaches a finite negative constant, so
\beq
\alpha_{IR,2\ell} \to 0^+ \quad {\rm as} \ \ N_f \nearrow N_{f,max} \ .
\label{alfir_2loop_upperend}
\eeq

For $N_f$ in the range where an IR zero of $\beta$ exists, it plays an
important role in the UV to IR evolution of the theory \cite{b2,bz}. If
$\alpha_{IR,2\ell}$ is large enough, then, as $\mu$ decreases through a scale
denoted $\Lambda$, the gauge interaction grows strong enough to produce a
bilinear fermion condensate in the most attractive channel, with attendant
spontaneous chiral symmetry breaking (S$\chi$SB) and dynamical generation of
effective masses for the fermions involved \cite{conf}. In a one-gluon exchange
approximation to the Dyson-Schwinger equation for the fermion propagator in a
vectorial gauge theory, this occurs as $\alpha$ increases through a value
$\alpha_{cr}$ given by $\alpha_{cr} C_f \sim O(1)$ \cite{wtc,chipt}.
Perturbative and nonperturbative corrections to this one-gluon exchange
approximation have been discussed \cite{dscor}.  In a chiral gauge theory this
fermion condensation breaks the gauge symmetry, while in the vectorial case,
the most attractive channel for fermion condensation is the singlet channel,
which preserves the gauge symmetry \cite{tc}.  Since the fermions that have
gained dynamical masses are integrated out in the low-energy effective field
theory below $\Lambda$, the $\beta$ function changes, and the theory flows away
from the original IR fixed point, which is thus only approximate.  However, if
$\alpha_{IR,2\ell}$ is sufficiently small, as is the case with a large enough
(AF-preserving) fermion content, then the theory evolves from the UV to the IR
without any spontaneous chiral symmetry breaking. In this case the theory has
an exact IR fixed point. 

 For a given $G$ and $N_f$ (massless) fermions in a representation $R$, the
critical value of $N_f$ beyond which the theory flows to the IR conformal phase
is denoted $N_{f,cr}$.  As $N_f$ increases, $\alpha_{IR,2\ell}$ decreases, and
$N_{f,cr}$ is the value at which $\alpha_{IR,2\ell}$ decreases through
$\alpha_{cr}$. The determination of the value of $N_{f,cr}$ for a given gauge
group $G$ and fermion representation $R$ is of basic field-theoretic interest.
In addition, this determination is important for ongoing studies of
quasi-conformal gauge theories.  These have a gauge coupling that gets large
but runs slowly over a long interval of $\mu$ due to an approximate IR fixed
point \cite{chipt}.  In the region of $N_f$ slightly less than $N_{f,cr}$,
where the theory confines but behaves in a quasi-conformal manner, some insight
has been gained from continuum studies of the changes in the spectrum of
gauge-singlet hadrons as compared with the spectrum in a QCD-like theory
\cite{chipt,bs1,bs2}. Going beyond continuum studies, there has been an
intensive recent program of lattice simulations to estimate $N_{f,cr}$ and
study the properties of quasi-conformal gauge theories. For example, recent
lattice papers on SU(3) with fermions in the fundamental representation include
\cite{lgt1,lgt2}, and some general reviews are given in the conferences
\cite{lgtrev}.  The UV to IR evolution of a chiral gauge theory and associated
sequential gauge symmetry breaking are also important for dynamical approaches
to fermion mass generation \cite{as}.

Since $\alpha_{IR,2\ell}$ is $\sim O(1)$, especially in the quasi-conformal
case where $N_f \lsim N_{f,cr}$, there are significant corrections to the
two-loop results from higher-loop terms in $\beta$.  These motivate one to
calculate these corrections to three-loop and four-loop order, and this has
been done in the $\overline{MS}$ scheme \cite{bvh,ps,bfs}. We found that, as
expected if perturbative calculations are reasonably reliable, for a given $R$
and $N_f$ (provided that $N_f \in I$, so that the two-loop beta function has an
IR zero), the shift in the location of the IR zero is smaller when one goes
from the three-loop to the four-loop level than when one goes from the two-loop
to the three-loop level.  The actual direction of the shift depends on the
fermion representation, $R$. For the fundamental (fund.) representation, we
found that, for a given $N$ and $N_f$,
\beq
\alpha_{IR,3\ell,\overline{MS}} < \alpha_{IR,4\ell,\overline{MS}} < 
\alpha_{IR,2\ell} \quad {\rm for} \ R=fund.
\label{alfir_loop_dep_msbar_fund}
\eeq
These shifts as a function of loop order are
larger for smaller $N_f$ and get smaller as $N_f$ approaches $N_{f,max}$. 
For example, for $G={\rm SU}(3)$ and $N_f=12$, we calculated
(cf. Table III of \cite{bvh})
\beqs
SU(3), N_f=12: \ \ & & \alpha_{IR,2\ell} = 0.754 \ , \cr\cr
                   & & \alpha_{IR,3\ell,\overline{MS}} = 0.435 \cr\cr
                   & & \alpha_{IR,4\ell,\overline{MS}} = 0.470 \ , \cr\cr
& & 
\label{alfir_nloopsu3nf12}
\eeqs
so the fractional shifts are 
\beqs
SU(3),  \ N_f=12: \ \ & &  
\frac{\alpha_{IR,3\ell,\overline{MS}}-\alpha_{IR,2\ell}}
     {\alpha_{IR,2\ell}} = -0.42 \cr\cr
& & 
\cr\cr
& & 
\frac{\alpha_{IR,4\ell,\overline{MS}}-\alpha_{IR,3\ell,\overline{MS}}}
     {\alpha_{IR,3\ell,\overline{MS}}} = +0.07 \ , 
\cr\cr
& & 
\label{su3nf12shifts}
\eeqs
and the resultant ratios are 
\beqs
SU(3), \ N_f=12: \ \ & &  
    \frac{\alpha_{IR,3\ell,\overline{MS}}}
         {\alpha_{IR,2\ell}} = 0.58 \cr\cr
& & 
\cr\cr
& & \frac{\alpha_{IR,4\ell,\overline{MS}}}
         {\alpha_{IR,2\ell}} = 0.62 \cr\cr
& & 
\cr\cr
& & \frac{\alpha_{IR,4\ell,\overline{MS}}}
         {\alpha_{IR,3\ell,\overline{MS}}}
= 1.08 \ . 
\cr\cr
& & 
\label{su3nf12ratios}
\eeqs
Qualitatively similar loop comparisons apply for other values of $N$ and $N_f$.

For the other (viz., adjoint and rank-2 symmetric and antisymmetric tensor)
representations that we studied in \cite{bvh}, we also found that higher-loop
values of the IR zero of $\beta$ were generically smaller than the two-loop
value, although not all parts of the inequality in
(\ref{alfir_loop_dep_msbar_fund}) necessarily held. As examples, for $G={\rm
SU}(2)$ and fermions in the adjoint (triplet) representation,
$\alpha_{IR,2\ell} = 0.628$, $\alpha_{IR,3\ell,\overline{MS}} = 0.459$, and
$\alpha_{IR,2\ell,\overline{MS}} = 0.450$, while for $G={\rm SU}(3)$ with octet
fermions, $\alpha_{IR,2\ell} = 0.419$, $\alpha_{IR,3\ell,\overline{MS}} =
0.306$, and $\alpha_{IR,2\ell,\overline{MS}} = 0.308$.

Clearly, it is important to assess the scheme-dependence in these calculations
of the IR zero of $\beta$ at three-loop and four-loop level. In particular, one
would like to know quantitatively how the value of the IR zero, computed at a
loop level higher than two loops, changes when one changes the scheme from the
$\overline{MS}$ scheme used in Refs. \cite{bvh,ps} to another scheme.  This
information is also useful for continuum studies of the boundary, as a function
of $N_f$ (for a given $N$ and $R$), between the IR phase with chiral symmetry
breaking and the chirally symmetric IR phase.  We address this question here.
First, we discuss general properties of a scheme transformation.


\section{Scheme Transformation}
\label{st}

\subsection{General} 
\label{generalst}

A scheme transformation can be expressed as a mapping between $\alpha$ and
$\alpha'$. It will be convenient to write this as
\beq
a = a' f(a') \ . 
\label{aap}
\eeq
To keep the UV properties the same, one requires $f(0) = 1$.  We consider 
that are analytic about $a=a'=0$ and hence can be expanded in the
form
\beq
f(a') = 1 + \sum_{s=1}^{s_{max}} k_s (a')^s =
        1 + \sum_{s=1}^{s_{max}} \bar k_s (\alpha')^s \ ,
\label{faprime}
\eeq
where the $k_s$ are constants, $\bar k_s = k_s/(4\pi)^s$, and $s_{max}$ may be
finite or infinite.  For $f(a')$ functions with infinite $s_{max}$, our
assumption of analyticity at $a'=a=0$ requires that the infinite series in 
Eq. (\ref{faprime}) converges within some nonzero radius of convergence. 
Given the form (\ref{faprime}), it follows that the Jacobian 
\beq
J = \frac{da}{da'}= \frac{d\alpha}{d\alpha'} 
\label{j}
\eeq
satisfies
\beq
J=1 \quad {\rm at } \ \ a=a'=0 \ . 
\label{jacobianazero}
\eeq
We have
\beq
\beta_{\alpha'} \equiv \frac{d\alpha'}{dt} = \frac{d\alpha'}{d\alpha} \, 
\frac{d\alpha}{dt} = J^{-1} \, \beta_{\alpha} \ . 
\label{betaap}
\eeq
This has the expansion 
\beq
\beta_{\alpha'} = -2\alpha' \sum_{\ell=1}^\infty b_\ell' (a')^\ell =
-2\alpha' \sum_{\ell=1}^\infty \bar b_\ell' (\alpha')^\ell \ ,
\label{betaprime}
\eeq
where $\bar b'_\ell = b'_\ell/(4\pi)^\ell$.  Given the equality of
Eqs. (\ref{betaap}) and (\ref{betaprime}), one can solve for the $b_\ell'$ in
terms of the $b_\ell$ and $k_s$. This leads to the well-known important 
result that \cite{gross75}
\beq
b_\ell' = b_\ell \quad {\rm for} \ \ \ell=1, \ 2 \ , 
\label{b12si}
\eeq
i.e., that the one- and two-loop terms in $\beta$ are scheme-independent.  We
note that the scheme-independence of $b_2$ assumes that $f(a')$ is
gauge-invariant. This is evident from the fact that in the momentum subtraction
(MOM) scheme, $b_2$ is actually gauge-dependent \cite{mom} and is not equal to
$b_2$ in the $\overline{MS}$ scheme.  We restrict our analysis here to
gauge-invariant scheme transformations and to schemes, such as $\overline{MS}$,
where $b_2$ is gauge-invariant.

If there is an IR zero in the two-loop $\beta_\alpha$, at
$\alpha_{IR,2\ell}$ given by (\ref{alfir_2loop}), then there is also an
IR zero in the two-loop $\beta_{\alpha'}$ at the same value of $\alpha'$,
This is consistent with the fact that, in general, (\ref{aap}) maps
$a'=-b_1/b_2$ to $a \ne -b_1/b_2$, since (\ref{aap}) is an exact
result, whereas the equality of two-loop IR zero values holds for the
truncations of $\beta_\alpha$ and $\beta_{\alpha'}$ to two-loop order.  This
difference is also important to remember in analyzing shifts of the location 
of the IR zero of $\beta$ function. For an illustration of this, we again take
$G={\rm SU}(3)$ and $N_f=12$. In Eqs. (\ref{alfir_nloopsu3nf12}) we listed the
values of $\alpha_{IR,2\ell}$ and, in the $\overline{MS}$ scheme, the values 
of the three-loop and four-loop IR zeros, $\alpha_{IR,n\ell,\overline{MS}}$,
$n=3,4$.  As an example of an acceptable scheme transformation, we consider the
application of the scheme transformation $a=(1/r)\sinh(ra')$ to the $\beta$
function in the $\overline{MS}$ scheme, which will be discussed in detail in
Section \ref{sinhst} below. For $r=6$, we find 
(cf. Table \ref{betazero_fund_sinh})
\beqs
S_{sh_r}, \ r=6: \quad & & \alpha'_{IR,2\ell,S_{sh_r};r=6}=
                           \alpha_{IR,2\ell} = 0.754 \cr\cr
                       & & \alpha'_{IR,3\ell,S_{sh_r};r=6}=0.433 \ , \cr\cr
                       & & \alpha'_{IR,4\ell,S_{sh_r};r=6}=0.467
\label{shvalues}
\eeqs
Because these zeros are calculated via truncations of the $\beta_{\alpha'}$
function to three-loop and four-loop order, respectively, they differ slightly
from the result of applying the exact (infinite-order) scheme transformation in
Eq. (\ref{sinhgen_inverse}) to the IR zeros in Eq. (\ref{alfir_nloopsu3nf12}).
Thus, the transformation $S_{sh_r}$ with $r=6$ maps the value 
$\alpha_{IR,2\ell} = 0.754$ to the value $0.739$, and so forth for the others
in Eq. (\ref{shvalues}).  In compact notation, 
\beqs
S_{sh_r;r=6}: \quad & & (\alpha_{IR,2\ell} = 0.754) \to 0.739 \ , \cr\cr
          & & (\alpha_{IR,3\ell,\overline{MS}} = 0.435) \to 0.432 \ , \cr\cr
          & & (\alpha_{IR,4\ell,\overline{MS}} = 0.470) \to 0.466 \ .
\label{sinhexact}
\eeqs
Similar comments apply for other values of $r$ with this $S_{sh_r}$ scheme
transformation, and for other scheme transformations. 
In general, for $N_f$ values where $\alpha_{IR,2\ell}$ is not too large, so
that the perturbative estimate of the IR zero of $\beta$ is reasonably
reliable, and provided that a scheme transformation is sufficiently 
well-behaved, the differences between $\alpha'_{IR,n\ell,S}$ calculated to
$n$-loop order and the result of applying the exact transformation to 
the initial scheme (here, the $\overline{MS}$ scheme) are small.  

For a given gauge group $G$ and fermion representation $R$, as $N_f$
approaches $N_{f,max}$ from below, since $\alpha_{IR,2\ell} \to 0$ as $N_f$
approaches $N_{f,max}$ from below (cf. Eq. (\ref{alfir_2loop_upperend})), it
follows that, insofar as higher-order perturbative calculations of $\beta$ are
reliable, they also yield $\alpha_{IR,n\ell} \to 0$ and, after an acceptable
scheme transformation, also
\beq
\alpha'_{IR,n\ell} \to 0^+ \quad {\rm as} \ \  N_f \nearrow N_{f,max} \ .
\label{alfirprime_upperend}
\eeq

In order to assess scheme-dependence of an IR fixed point, we have calculated
the relations between the $b'_\ell$ and $b_\ell$ for higher $\ell$. For
example, for $\ell=3, \ 4, \ 5$ we obtain
\beq
b_3' = b_3 + k_1b_2+(k_1^2-k_2)b_1 \ , 
\label{b3prime}
\eeq
\beq
b_4' = b_4 + 2k_1b_3+k_1^2b_2+(-2k_1^3+4k_1k_2-2k_3)b_1 \ , 
\label{b4prime}
\eeq
and
\beqs
b_5' & = & b_5+3k_1b_4+(2k_1^2+k_2)b_3+(-k_1^3+3k_1k_2-k_3)b_2 \cr\cr
     & + & (4k_1^4-11k_1^2k_2+6k_1k_3+4k_2^2-3k_4)b_1 \ .
\label{b5prime}
\eeqs
We list the somewhat longer expressions for $b'_\ell$ for $\ell=6, \ 7, \ 8$
in the Appendix. Since the $b_\ell$ have been calculated only up to $\ell=4$,
we will only need the above results for $b'_3$ and $b'_4$ in our study of
the effect of performing scheme transformations on the four-loop $\beta$
functions for a non-Abelian gauge theory.  However, we will use the $b'_\ell$
up to $\ell=8$ in our study of the effect of scheme transformations on an
illustrative hypothetical exact $\beta$ function in Section \ref{exactbeta}. 

From the expressions for $b'_\ell$ with $3 \le \ell \le 8$ that we have
calculated, we can discern several general structural properties. First, in the
coefficients of the terms $b_n$ entering in the expression for $b_\ell$, the
sum of the subscripts of the $k_s$s is equal to $\ell-n$ with $1 \le n \le
\ell-1$, and the products of the various $k_s$s correspond to certain
partitions of $\ell-n$. For example, in the expression for $b_4'$, the
coefficient of $b_1$ contains the term $-2k_1^3$ corresponding to the partition
(1,1,1) of $\ell-n=4-1=3$, the term $4k_1k_2$ corresponding to the partition
(1,2) of 3, and the term $-2k_3$, corresponding to the partition 3 of
3. However, because of cancellations, in the expression for $b_\ell'$ for even
$\ell$, the coefficient of $b_n$ does not contain all of the terms
corresponding to the partitions of $\ell-n$.  For example, in the expression
for $b_2'$, there is no $k_1b_1$ term; in the expression for $b_4'$, although
the partitions of 2 are $\{(1,1),(2)\}$, the coefficient of $b_2$ does not
contain $k_2$; and in the expression for $b_6'$, although the partitions of 3
are $\{(1,1,1), (1,2), (3) \}$, the coefficient of $b_3$ does not contain
$k_1^3$ or $k_3$.  A corollary of the structural property above is that the
only $k_s$s that appear in the formula for $b_\ell'$ are the $k_s$s with $1 \le
s \le \ell-1$.

We note that the form for $f(a')$ in Eq. (\ref{faprime}) could be generalized
further so that $f(a')$ could include a part that is 
finite but nonanalytic at $a'=0$. An example is 
\beq
f(a') = [1+\sum_{s=1}^{s_{max}} k_s (a')^s ][1 + \kappa e^{-\nu/a'}] \ , 
\label{faprimenonanalytic}
\eeq
where $\kappa$ and $\nu$ are (real) constants and $\nu > 0$.  (In this context,
we recall that expressions containing terms like $\exp(-8\pi^2/g^2)$
naturally arise in instanton calculations.)  Since no terms involving $\kappa$
occur at any finite order of a perturbative expansion of $f(a')$ in powers of
$a'$, our results for $b'_\ell$ in Eqs. (\ref{b3prime})-(\ref{b5prime}),
(\ref{b6prime})-(\ref{b8prime}) continue to hold for these scheme
transformations.


\subsection{Transformation to 't Hooft Scheme at a UVFP}
\label{thooftscheme}

Given that the $b_\ell$ for $\ell \ge 3$ are scheme-dependent, one may ask
whether it is possible to transform to a scheme in which the $b'_\ell$ are all
zero for $\ell \ge 3$, i.e., a scheme in which the two-loop $\beta$ function is
exact.  Here and elsewhere, by the term ``exact two-loop $\beta$ function'' we
mean exact in the sense of Eq. (\ref{beta}), which does not include possible
nonperturbative contributions, such as could be produced by instantons
\cite{cdg}.  Near the UV fixed point at $\alpha=0$, this is possible, as
emphasized by 't Hooft \cite{thooft77}.  This is commonly called the 't Hooft
scheme, and we denote it as $S_H$. For this and other schemes, we shall also
use this symbol to refer to transformation that takes one to the given target
scheme; the meaning will be clear from the context.

We next present an explicit scheme transformation which, starting from a given
scheme, transforms to the 't Hooft scheme. This necessarily has
$s_{max}=\infty$.  Our key to constructing this transformation is to take
advantage of the property that $b'_\ell$ for $\ell \ge 3$ contains only a
linear term in $k_{\ell-1}$, so that the equation $b'_\ell=0$ is a linear
equation for $k_{\ell-1}$, which can always be solved.  In order to simplify
the transformation, we start by setting $k_1=0$.  We then solve the equation
$b'_3=0$ for $k_2$, obtaining
\beq
k_2 = \frac{b_3}{b_1} \ . 
\label{k2thooftsol}
\eeq
We then substitute these values of $k_1$ and $k_2$ into the equation $b'_4=0$
using our expression (\ref{b4prime}) and solve for $k_3$, obtaining
\beq
k_3 = \frac{b_4}{2b_1} \ . 
\label{k3thooftsol}
\eeq
We then continue iteratively in this manner.  In the next step, we substitute
these values of $k_s$, $s=1,2,3$, into the expression for $b'_5=0$, using
Eq. (\ref{b5prime}), and solve for $k_4$, getting 
\beq
k_4 = \frac{b_5}{3b_1} - \frac{b_2b_4}{6b_1^2} + \frac{5b_3^2}{3b_1^2} \ . 
\label{k4thooftsol}
\eeq
Proceeding in this manner, we obtain
\beq
k_5 = \frac{b_6}{4b_1} - \frac{b_2b_5}{6b_1^2} + \frac{2b_3b_4}{b_1^2} 
+ \frac{b_2^2b_4}{12b_1^3} - \frac{b_2b_3^2}{12b_1^3} 
\label{k5thooftsol}
\eeq
and
\beqs
k_6 &  = & \frac{b_7}{5b_1} - \frac{3b_2b_6}{20b_1^2} + \frac{8b_3b_5}{5b_1^2} 
+ \frac{11b_4^2}{20b_1^2} \cr\cr
& - & \frac{4b_2b_3b_4}{5b_1^3} + \frac{b_2^2b_5}{10b_1^3} + 
\frac{16b_3^3}{5b_1^3} + \frac{b_2^2b_3^2}{20b_1^4}-\frac{b_2^3b_4}{20b_1^4} \
.
\cr\cr
& & 
\label{k6thooftsol}
\eeqs
One can continue this procedure iteratively to calculate $k_s$ with arbitrarily
high values in $s$, since the equation $b_\ell=0$ is a linear equation for
$k_{\ell-1}$, which always has a solution.  This yields a two-loop $\beta$
function that is exact.  We shall refer to this transformation as the $S_H$
transformation.  Although we do not claim that this is the only way to
transform to the 't Hooft scheme, it is a particularly simple way to do so.

There are several salient structural features of these expressions for the
$k_s$s.  First, $k_s$ only depends on ratios of $b_\ell/b_1$. Second, the
$\ell$ values that occur in the ratios $b_\ell/b_1$ that enter into the
expression for $k_s$ have the property that in a term proportional to
\beq
\frac{ \prod_{i=2}^{i_{max}} b_{\ell_i}}{b_1^j} \ , 
\label{kstermthooft}
\eeq
one has 
\beq
s = \Big [ \sum_{i=2}^{i_{max}} \ell_i \Big ] - j \ , 
\label{strel}
\eeq
where $i_{max}$ is determined by $s+j$.  As a corollary, the sets of $\ell_i$
that enter into the numerator of Eq. (\ref{kstermthooft}) arise as subsets of
the partitions of $s+j$ that exclude the integer 1.  For example, in the
expression for $k_6$, Eq. (\ref{k6thooftsol}), the products of $b_{\ell_i}$
that enter in the terms proportional to $b_1^{-2}$ have sets of $\ell_i$ values
that are a subset of partitions of $6+2=8$ that exclude the value 1, including
(2,6), (3,5), and (4,4), corresponding to the products $b_2b_6$, $b_3b_5$, and
$b_4^2$. Not all of the partitions of $s+j$ excluding 1 are represented; in the
example given, the partitions of 8 excluding 1 also include (2,2,2,2), (2,2,4),
and (2,3,3), but the numerators of these terms proportional to $b_1^{-2}$ in
$k_6$ do not include $b_2^4$, $b_2^2b_4$, or $b_2b_3^2$.

In this 't Hooft scheme with a (perturbatively) exact two-loop $\beta$
function, if the resultant IR zero, $\alpha_{IR,2\ell}$, is at a sufficiently
small coupling to lie in the non-Abelian Coulomb phase so that the evolution
into the infrared does not entail any spontaneous chiral symmetry breaking or
attendant dynamical fermion mass generation, then this is an exact IR fixed
point.  In this case, one can take advantage of the exact solution of the
differential equation represented by the two-loop $\beta$ function in terms of
a Lambert function \cite{gardi}.  In contrast, if the resultant IR zero,
$\alpha_{IR,2\ell}$, is greater than the critical value, $\alpha_{cr}$ for
spontaneous chiral symmetry breaking and associated bilinear fermion condensate
formation, then, as $\mu$ decreases below a scale denoted $\Lambda$ and
$\alpha$ increases past $\alpha_{cr}$, this condensate formation occurs, the
fermions gain dynamical masses, and one integrates them out of the low-energy
effective field theory applicable below this scale.  Hence, the $\beta$
function changes to that of a pure gluonic theory, and so one cannot use the
solution in terms of a Lambert function calculated for $\mu > \Lambda$, but
instead must match this onto a different solution with $N_f=0$ applicable for
$\mu < \Lambda$. This latter solution does not involve any perturbative IR
zero.


\subsection{Necessary Conditions for an Acceptable Scheme Transformation}
\label{conditions}

In order to be physically acceptable, this transformation must satisfy several
conditions, $C_i$.  For finite $s_{max}$, Eq. (\ref{aap}) is an algebraic
equation of degree $s_{max}+1$ for $\alpha'$ in terms of $\alpha$.  We require
that at least one of the $s_{max}+1$ roots must satisfy these conditions.  For
$s_{max}=\infty$ with nonzero $k_s$ for arbitrarily large $s$, the equation for
$\alpha'$ in terms of $\alpha$ is generically transcendental, and again we
require that the physically relevant solution must satisfy these conditions.
These are as follows:
\begin{itemize}

\item 

$C_1$: the scheme transformation must map a real positive $\alpha$ to a real
positive $\alpha'$, since a map taking $\alpha > 0$ to $\alpha'=0$ would be
singular, and a map taking $\alpha > 0$ to a negative or complex $\alpha'$
would violate the unitarity of the theory.

\item  

$C_2$: the scheme transformation should not map a
moderate value of $\alpha$, for which perturbation theory may be reliable, to a
value of $\alpha'$ that is so large that perturbation theory is
unreliable. 

\item 

$C_3$: \ $J$ should not vanish in the region of $\alpha$ and
$\alpha'$ of interest, or else there would be a pole in Eq. (\ref{betaap}).

\item

$C_4$: \quad The existence of an IR zero of $\beta$ is a scheme-independent
property of an AF theory, depending (insofar as perturbation theory is
reliable) only on the condition that $b_2 < 0$.  Hence, a scheme transformation
must satisfy the condition that $\beta_\alpha$ has an IR zero if and only if
$\beta_{\alpha'}$ has an IR zero.

\end{itemize}

Since one can define a transformation from $\alpha$ to $\alpha'$ and the
inverse from $\alpha'$ to $\alpha$, these conditions apply going in both
directions.  These four conditions can always be satisfied by scheme
transformations used to study the UV fixed point and hence in applications to
perturbative QCD calculations, since the gauge coupling is small (e.g.,
$\alpha_s(m_Z) = 0.118$), and one can choose the $k_s$ to have appropriately
small magnitudes.  By continuity, it follows that among the $s_{max}+1$ roots
of Eq. (\ref{aap}), there is always one with a real (positive) $\alpha' \simeq
\alpha$ near the UV fixed point at $\alpha=0$.  For small $\alpha$, $C_1$-$C_4$
are then met.  We note that, in addition to these four conditions, there may
also be other related ones that must be satisfied for a given scheme
transformation to be acceptable. For example, in the $S_1$ scheme presented in
\cite{scc}, it is necessary that the expression $b_2^2-4b_1b_3$ in
Eq. (\ref{k1pm}) must be nonnegative.


\section{Examples of Scheme Transformations Acceptable at a UVFP
  but not at an IRFP} 
\label{pathologicalschemes}

In \cite{scc} we pointed out that although these conditions $C_1$-$C_4$ can
easily be satisfied by a scheme transformation applied in the vicinity of the
UV fixed point at $\alpha=\alpha'=0$, they are not automatically satisfied, and
are a significant constraint, on a scheme transformation that one tries to
apply in the vicinity of an IR fixed point. In \cite{scc} we demonstrated this
with two specific examples: (i) $\alpha = \alpha' \tanh(\alpha')$, and (ii) a
scheme transformation with $s_{max}=2$, $k_1=0$, and $k_2=b_3/b_1$ designed to
render $b'_3=0$.  Here we elaborate on these, give a third example of a scheme
transformation that is acceptable at a UV fixed point but not at a general IR
fixed point, and discuss some issues that arise with a fourth transformation.
The two pathological transformations presented in \cite{scc} are denoted,
respectively, as (i) the special $r=4\pi$ case of the $S_{th_r}$ scheme
transformation and (ii) the $S_2$ scheme transformation, discussed below in
Sections \ref{tanhst} and \ref{s2st}, respectively.  Our two additional
examples are the $S_3$ scheme transformation in Section {\ref{s3st} and the
transformation $S_H$ to the 't Hooft scheme in Section \ref{sthooft}.  In the
following, to avoid overly complicated notation, we will use the generic
notation $\alpha'$ for the result of the application of each scheme
transformation to an initial $\alpha$, with it being understood that this
refers to the specific transformation under consideration. Where it is
necessary for clarify, we will use a subscript to identify the specific scheme
$S$ being discussed.


\subsection{The $S_2$ Transformation with $s_{max}=2$ to a Scheme with 
$b'_3=0$}
\label{s2st}

Here we elaborate on the scheme transformation discussed in \cite{scc} with
$s_{max}=2$ that renders $b'_3=0$ and is acceptable at a UV fixed point, but
was shown to be unacceptable at a general IR fixed point.  Because $s_{max}=2$,
this scheme transformation depends on two parameters, $k_s$ with $s=1,2$. Since
$b'_3$ depends quadratically on $k_1$ and linearly on $k_2$, the solution of
the desired condition $b'_3=0$ is simplest if one sets $k_1=0$.  Then, using
Eq. (\ref{b3prime}) and solving this equation $b'_3=0$ for $k_2$, one finds
\beq
k_2 = \frac{b_3}{b_1} \ .
\label{k2st2}
\eeq
This scheme transformation, denoted $S_2$, is then 
\beqs
S_2: \quad\quad & & s_{max} = 2, \quad k_1 = 0 \ , \quad k_2=b_3/b_1 \ , 
i.e., \cr\cr
     & & a=a' \Big [ 1 + \frac{b_3}{b_1}(a')^2 \Big ] \ .
\label{s2}
\eeqs
Applying this $S_2$ scheme transformation to an initial scheme, one obtains
\beq
b'_4=b_4 \ . 
\label{bp4s2}
\eeq
It is straightforward to calculate the $b'_\ell$ for $\ell \ge 5$, but we 
will not need them here. 

By construction, at the three-loop level, $\beta_{\alpha'}$ in this scheme is
the same as the (scheme-independent) two-loop $\beta$ function, so the IR zero
of $\beta_{\alpha'}$ at the three-loop level is
\beqs
\alpha'_{IR,3\ell,S_i} & = & \alpha'_{IR,2\ell,S_i} = \alpha_{IR,2\ell} = 
-\frac{4\pi b_1}{b_2} \cr\cr
& & {\rm for} \ \ S_i=S_1, \ S_2, \ S_3. 
\label{s2ttrel}
\eeqs
(We write this in a general form, since it holds not just for the present $S_2$
scheme transformation, but also for the $S_3$ and $S_1$ transformations to be
discussed below.)  At the four-loop level in this $S_2$ scheme, the IR zero is
determined by the physical (smallest positive) solution of the cubic equation
\beq
b_1 + b_2 a' + b_4' (a')^3 = 0 \ . 
\label{cubiceqs2}
\eeq

In order that this transformation obey condition $C_1$, that it maps $a' > 0$
to $a > 0$, we require that $1+(b_3/b_1)(a')^2 > 0$. This inequality 
must be satisfied, in particular, in
the vicinity of the two-loop IR zero of $\beta$, so substituting the
(scheme-independent) $a_{IR,2\ell}=a'_{IR,2\ell}=-b_1/b_2$ from Eq.
(\ref{alfir_2loop}), we obtain the inequality
\beq
1 + \frac{b_1b_3}{b_2^2} > 0 \ .
\label{conditionc2irst2}
\eeq
But, as noted in \cite{scc}, this inequality is not, in general, satisfied. 
For example, let us consider the class of theories with
$G={\rm SU}(N)$ and $N_f$ fermions in the fundamental representation.
Substituting the scheme-independent expressions for $b_1$ and $b_2$
\cite{b1,b2}, together with the expression for $b_3$ in the $\overline{MS}$
scheme \cite{b3} for this class of theories, the inequality
(\ref{conditionc2irst2}) becomes
\begin{widetext}
\beqs
& & \frac{ 104470N^6 + 3N_fN(-26950N^4+4505N^2+99)+N_f^2(15384N^4-4656N^2+270)
+4N_f^3N(-112N^2+33) }{36 [34N^3+N_f(-13N^2+3) ]^2 } > 0 \ . 
\cr\cr
& & 
\label{conditionc2irst2fund}
\eeqs
\end{widetext}
For a given value of $N$, the determination of the range in $N_f$ where this
inequality is satisfied involves the calculation of the zeros of the numerator
of (\ref{conditionc2irst2fund}), which are solutions of a cubic equation in
$N_f$. For $N=2$, these zeros occur at $N_f = 4.27, \ 8.44, \ 55.90$, while for
$N=3$ they occur at $N_f=6.22, \ 12.41, \ 84.32$. As before, we restrict our
consideration to the interval $I$ given by Eq. (\ref{nfinterval}), $N_{f,b2z} <
N_f < N_{f,max}$, where the two-loop $\beta$ function has an IR zero.  For
$N=2$, this interval $I$ is $5.55 < N_f < 11$, and in this interval the
inequality is violated for $5.55 < N_f < 8.44$ and is satisfied for $8.44 < N_f
< 11$.  For $N=3$, the interval $I$ is $8.05 < N_f < 16.5$, and in this
interval, the inequality is violated for $8.05 < N_f < 12.41$ and is satisfied
for $12.41 < N_f < 16.5$.  For the physical, integer values of $N_f$, these
statements are evident from the values of $\bar b_\ell$ listed in Table III of
our Ref. \cite{bvh}.  For example, for $N=3$ and $N_f=10$, where
$\alpha_{IR,2\ell}=2.21$, the values of these coefficients are $\bar b_1=
0.345$, $\bar b_2=-0.156$, and $\bar b_3=-0.386$, so that
\beqs
& & 1+\frac{\bar b_1 \bar b_3}{\bar b_2^2}  = 1 + \frac{b_1 b_3}{b_2^2} = 
-4.47 \cr\cr
& & {\rm for} \ \ G = {\rm SU}(3), \ N_f=10, \ R=fund. \ .
\label{ineqfornc3nf10}
\eeqs
The values of $1+(b_1b_3/b_2^2)$ for $N=3$ and some larger values of $N_f$ are
as follows: $-1.43$ for $N_f=11$ and $-0.270$ for $N_f=12$, with positive
values for $N_f \ge 13$ in the interval $I$, including the value $+0.293$ for
$N_f=13$. 

The pathology that this $S_2$ scheme transformation violates conditions $C_1$
and $C_4$ is reflected in the results that one gets by actually applying it to
the four-loop $\beta$ function in the $\overline{MS}$ scheme and solving for
the IR zeros.  As above, we focus on the case of fermions in the fundamental
representation, with $N_f \in I$.  We list the values of
$\alpha'_{IR,3\ell,S_2}$ and $\alpha'_{IR,4\ell,S_2}$ in Table
\ref{betazero_fund_s1s3}.  The three-loop values are given by Eq.
(\ref{s2ttrel}). As regards the four-loop values, we find that,
except for $N_f$ value(s) near $N_{f,max}$, at the upper end of the non-Abelian
Coulomb phase, the cubic equation (\ref{cubiceqs2}) yields a negative root and
a complex-conjugate pair of roots, none of which is physically acceptable.  For
example, for $N=2$, there is no physical root (denoted as n.p. in the table)
for $N_f \in I$ except for the highest value of $N_f$ below $N_{f,max}$, namely
$N_f=10$. Similarly, when $N=3$, a physical root of the cubic equation first
appears for $N_f=14$ and when $N=4$, this happens when $N_f=19$.

Thus, although this $S_2$ scheme transformation is acceptable at the UV fixed
point at $\alpha=0$ and at a sufficiently weakly coupled IR fixed point at the
upper end of the non-Abelian Coulomb phase, it is not acceptable at a general
IR fixed point, since it fails to satisfy condition $C_1$. The latter pathology
occurs when $\alpha_{IR}$ grows to a value of order unity.  According to the
results of several lattice groups \cite{lgt1}, for $N=3$, the theory with
$N_f=12$, and hence also the theory with $N_f=13$, evolve into the infrared in
a conformal, non-Abelian Coulomb phase (other lattice groups differ on the
$N_f=12$ case \cite{lgt2}).  Provided that one accepts that $N_f=12$, and hence
also $N_f=13$, are in the non-Abelian Coulomb phase, our results above show
that a scheme transformation may fail to be acceptable not only at an IR fixed
point in the phase with confinement and spontaneous chiral symmetry breaking
(which is approximate), but also at an exact IR fixed point in the lower part
of the chirally symmetric conformal phase.


\subsection{The $S_3$ Transformation with $s_{max}=2$ to a Scheme with 
$b'_3=0$}
\label{s3st}

Here we present a scheme transformation with $s_{max}=2$ that is also designed
to render $b'_3=0$ and is acceptable at a UV fixed point, but we show that it
is not acceptable at a general IR fixed point.  The property that $s_{max}=2$
and the goal of rendering $b'_3=0$ are the same as those of the transformation
given in \cite{scc} (denoted $S_2$ here). Since one uses a scheme
transformation with $s_{max}=2$ and since $b'_3$ depends only on $k_1$ and
$k_2$, it follows that a natural first choice is to try $k_1=0$ and $k_2 \ne
0$.  This was the ($S_2$) transformation that was shown to be unacceptable at a
general IR fixed point in \cite{scc}. Another natural choice is to set
$k_2=k_1^2$, since this renders the coefficient of the $b_1$ term in $b'_3$,
namely $(k_1^2-k_2)$, equal to zero. Hence, this choice considerably simplifies
the equation $b'_3=0$, which is reduced to a linear equation for $k_1$, with
solution $k_1 = -b_3/b_2$.  We denote this scheme transformation as $S_3$,
\beq
S_3: \quad s_{max}=2, \quad k_1 = -\frac{b_3}{b_2} \ , \quad 
k_2 = k_1^2 = \frac{b_3^2}{b_2^2}
\label{s3}
\eeq
and study it here.  As before, we also use $S_3$ to refer to the scheme that is
obtained by applying this transformation to an initial scheme such as the
$\overline{MS}$ scheme.  We denote the resultant IR zero of $\beta_{\alpha'}$
at the $n$-loop level as $\alpha'_{IR,n\ell,S_3}$.  Evaluating
Eq. (\ref{b4prime}) for $b'_4$ in this scheme, we calculate
\beq
b'_4 = b_4 - \frac{b_3^2}{b_2} - \frac{2b_1 b_3^3}{b_2^3} \quad {\rm for} \ \ 
S_3 \ . 
\label{bp4s3}
\eeq
The function $f(a')$ takes the simple form
\beqs
f(a') & = & 1+\xi+\xi^2 \quad {\rm for} \ S_2 \ , \cr\cr
      & & {\rm where} \ \ \xi \equiv k_1a' = -\frac{b_3a'}{b_2} \ . 
\label{faprimes3}
\eeqs
Now $1+\xi+\xi^2$ is always positive, with no real zero in $\xi$ (and a minimum
at $\xi=-1/2$, where this polynomial is equal to 3/4). The 
Jacobian for this transformation is
\beq
J=1+2\xi+3\xi^2  \quad {\rm for} \ \ S_3 . 
\label{js2}
\eeq
This $J$ is also positive, with no real zero in $\xi$ (and a minimum at
$\xi=-1/3$, where $J=2/3$).  As with the $S_2$ scheme, at the three-loop level,
$\beta_{\alpha'}$ in this scheme is the same as the two-loop $\beta$ function,
so the IR zero of $\beta_{\alpha'}$ at the three-loop level satisfies
Eq. (\ref{s2ttrel}).  At the four-loop level in this
$S_3$ scheme, the IR zero is determined by the physical (smallest positive)
solution of the cubic equation (\ref{cubiceqs2}) with $b'_4$ given by
Eq. (\ref{bp4s3}).

We have calculated the resultant $\alpha'_{IR,n\ell} \equiv
\alpha_{IR,n\ell,S_3}$ in this $S_3$ scheme up to the $(n=4)$-loop level. In
Table \ref{betazero_fund_s1s3} we list values of the $n$-loop IR zero,
$\alpha'_{IR,n\ell,S_3}$ for $n=2,3,4$ for relevant $N_f$, with fermions in the
fundamental representation and several values of $N$.  For comparison we also
include the values of $\alpha_{IR,n\ell,\overline{MS}}$ for $n=3,4$ in the
$\overline{MS}$ scheme from \cite{bvh}. Since the two-loop value is
scheme-independent, we denote it simply as $\alpha_{IR,2\ell}$. The relation
(\ref{s2ttrel}) is reflected in the entries in the table. The
four-loop zero is denoted as $\alpha'_{IR,4\ell,S_3}$.  In contrast with
$\alpha_{IR,n\ell,\overline{MS}}$ and $\alpha'_{IR,n\ell,S_1}$, which decrease
monotonically as a function of $N_f$ for a given $N$, $\alpha'_{IR,4\ell,S_3}$
behaves nonmonotonically as a function of $N_f$, first increasing and then
decreasing.

But our overriding result here is that the $S_3$ scheme transformation does not
yield any physical value for $\alpha'_{IR,4\ell,S_3}$ in the case of SU(4) with
$N_f=18$ in the fundamental representation. In this case, the above-mentioned
cubic equation has only a negative root and a complex-conjugate pair of roots.
Hence, this $S_3$ scheme transformation fails conditions $C_1$ and $C_4$ and
must be rejected as unacceptable in the vicinity of a general IR fixed point.
This theory, with an SU(4) gauge group and $N_f=18$ fermions is likely to be in
a non-Abelian Coulomb phase in the infrared.  Assuming this is the case, this
provides another example of how a scheme transformation can be
pathological not just in the confined phase with spontaneous chiral symmetry
breaking, but also in the infrared conformal phase. 


\subsection{The $S_H$ Transformation to the 't Hooft Scheme} 
\label{sthooft}

In Section \ref{thooftscheme} we have constructed a scheme transformation that
can be applied to an arbitrary initial scheme to shift to the 't Hooft scheme,
with $b'_\ell=0$ for $\ell \ge 3$ and thus a (perturbatively) exact two-loop
$\beta$ function.  By the general continuity arguments that we have presented,
this scheme transformation satisfies all of the requisite conditions to be an
acceptable transformation in the vicinity of the UV fixed point at
$\alpha=\alpha'=0$. However, one encounters a complication with this
transformation at an IR fixed point.  This can be explained as
follows. For a given group $G$ and fermion representation $R$, as $N_f$
increases toward $N_{f,max}$, $b_1 \to 0$, while $b_2$ and, in the initial
scheme, the $b_\ell$ with $\ell \ge 3$, approach finite nonzero values.  Hence,
since the coefficient $k_s$ is a sum of terms each of which contains an inverse
power of $b_1$, it follows that, as $N_f$ takes on values close to $N_{f,max}$,
these $k_s$ coefficients may have arbitrarily large magnitudes as $s \to
\infty$. For a particular term $k_s (a')^s$ in the sum (\ref{faprime}), much of
this growth is cancelled, since, $a'_{IR,2\ell} \propto b_1$.  However, since
one must use an infinite number of $k_s$ terms to render all of the $b'_\ell$
equal to zero for this $S_H$ transformation, one encounters the issue of the
convergence of the infinite series for $f(a')$ in Eq. (\ref{faprime}).  Note
that this is not an issue of strong coupling, as are the pathologies in the
$S_2$, $S_3$, and $S_{th_r}$ scheme transformations; it occurs in the weakly
coupled, non-Abelian Coulomb phase.  We do not claim here that it is impossible
to construct an acceptable scheme transformation to get to the 't Hooft scheme
in the vicinity of an IR fixed point, only that one encounters delicate issues
of convergence with the $S_H$ scheme, since for a fixed $N_f$ near to
$N_{f,max}$, the $k_s$ may have unbounded magnitudes as $s \to \infty$.

As we will discuss below, the scheme transformation $S_1$ contains a parameter
(denoted $k_{1p}$) that also grows large as $N_f$ approaches $N_{f,max}$, but,
although inconvenient, this is much less serious, since there is only a single
parameter involved, since $s_{max}=1$, not an infinite number, as with the
$S_H$ transformation, and the growth of this single parameter, restricted to
integer values of $N_f$, is bounded.


\section{The Transformation $S_1$ with $s_{max}=1$ to a Scheme with $b'_3=0$}
\label{s1}

We next proceed to construct and study scheme transformations that are
acceptable at an (exact or approximate) IR fixed point and use them to study
the scheme dependence of the location of this fixed point.  For comparative
purposes, it is useful to begin by discussing the scheme denoted $S_1$ that we
presented in \cite{scc}, on which we will give more details here. 

The original motivation for our construction of this $S_1$ scheme
transformation was the idea of designing a transformation that would render at
least one of the $b'_\ell$ with $\ell \ge 3$ equal to zero, namely $b'_3$. In
turn, this was motivated by the idea of having a scheme transformation that
achieves at least one step in the sequence of steps that defines a
transformation to the 't Hooft scheme, where $b'_\ell=0$ for all $\ell \ge 3$.
The next steps in this direction would be design a scheme transformation that
would render both $b'_3=0$ and $b'_4=0$ at an IR fixed point, and then one that
would render $b'_\ell =0$ for $\ell=3,4,5$, and so forth, up to a fixed value
of $s$.  As a reasonable first exploration of such endeavors, we opted to focus
on scheme transformations that rendered just $b'_3=0$.  We have considered
three of these, labelled $S_j$, $j=1,2,3$, and shown that the $S_2$ and $S_3$
transformations are not acceptable at a general IR fixed point.  As we will
show below, the $S_1$ scheme transformation has the inconvenient feature that
the $k_s$ coefficients grow as one approaches the upper end of the non-Abelian
Coulomb phase, producing a rather strong scheme-dependence even at the
four-loop level.  This $S_1$ scheme transformation is, nevertheless, valuable
as a lesson that shows how large scheme-dependent effects can be.  As we will
show below in Section \ref{sinhgen}, the $S_{sh_r}$ scheme transformation in
Eq. (\ref{sinhgen}) with moderate values of $r$ is better-behaved and, when
applied to the $\beta$ function in the $\overline{MS}$ scheme, produces smaller
shifts in the location of the IR zero than the $S_1$ transformation.

We proceed to the details of the construction of the $S_1$ scheme
transformation presented in \cite{scc}.  We assume $N_f \in I$, so a two-loop
IR zero of $\beta$ exists.  Since $s_{max}=1$, Eq. (\ref{aap}) reads
$a=a'(1+k_1 a')$.  Although this quadratic equation has two formal solutions,
only the solution
\beq
\alpha'_+ = \frac{1}{2\bar k_1} \Big (-1 + \sqrt{1+4 \bar k_1 \alpha} \ 
\Big ) 
\label{alfprimesmax1}
\eeq
is acceptable, since only this solution has $\alpha \to \alpha'$ as $\alpha \to
0$.  

This scheme transformation was designed to render $b'_3=0$, so the
next step is to solve the equation $b'_3=0$ using Eq. (\ref{b3prime}), viz.,
\beq
b_3+k_1b_2+k_1^2b_1=0 \ , 
\label{b3ps1}
\eeq
for the parameter $k_1$.  Formally, Eq. (\ref{b3ps1}) has two solutions,
\beq
k_{1p}, \ k_{1m} = \frac{1}{2b_1} \Big [ -b_2 \pm \sqrt{b_2^2-4b_1b_3} \
\Big ] \ ,
\label{k1pm}
\eeq
where $(p,m)$ refer to $\pm$.  We will focus on $G={\rm SU}(N)$ with fermions
in the fundamental and adjoint representation. Of the two formal solutions in
Eq. (\ref{k1pm}), only $k_{1p}$ is allowed.  To show this, we consider
$k_{1m}$.  We must be able to use this for $N_f \in I$, including the lower end
of this interval, where $N_f$ approaches $N_{f,b2z}$ from above.  Precisely at
the lower end, as $N_f \searrow N_{f,b2z}$, $b_2 \to 0^-$ and
$\alpha_{IR,2\ell} \to \infty$, so clearly one cannot trust perturbative
calculations at or near this point.  However, we will at least require that the
transformation should obey the conditions $C_1$-$C_4$ for $N_f \gsim N_{f,b2z}$
where $\alpha_{IR,2\ell}$ is not too large.  As shown in \cite{bvh}, in this
region of $N_f$, $b_3 < 0$, so that, taking into account that both $b_2$ and
$b_3$ are negative in this region, we can reexpress $k_{1m}$ as
\beq
k_{1m} = \frac{1}{2b_1} \Big [ |b_2| - \sqrt{b_2^2+4b_1|b_3|} \ \Big ] \quad 
{\rm for } \ N_f \gsim N_{f,b2z} \ .
\label{km1form}
\eeq
As $N_f \searrow N_{f,b2z}$, $b_2 \to 0^-$, so $k_{1m} \to
-\sqrt{|b_3|/b_1}$.  Substituting this into Eq. (\ref{alfprimesmax1}), using
$\bar k_1=k_1/(4\pi)$, we have 
\beqs
\alpha'_+ &=& \frac{1}{2\bar k_{1m}}\Big (-1 + \sqrt{1+4 \bar k_{1m} \alpha} 
\ \Big ) 
\cr\cr    &=& \frac{1}{2|\bar k_{1m}|}\Big ( 1 - \sqrt{1-4|\bar k_{1m}|\alpha}
\ \Big ) \ . 
\label{alfprimesmax1form}
\eeqs
Next, substituting the value of $\alpha_{IR,2\ell}$ from
Eq. (\ref{alfir_2loop}) as a relevant estimate, the square root in
Eq. (\ref{alfprimesmax1}) becomes
\beq
\bigg [ 1 - \frac{\sqrt{b_1|b_3|}}{|b_2|} \ \bigg ]^{1/2} \ . 
\label{k1maux}
\eeq
As $N_f$ approaches $N_{f,b2z}$ from above and $b_2 \to 0^-$, the expression in
this square root becomes negative, so that the square root itself is imaginary.
Hence, if one were to try to use $k_{1m}$ with this scheme transformation, then
a real $\alpha \simeq \alpha_{IR,2\ell}$ would get mapped via
Eq. (\ref{alfprimesmax1}) to a complex, unphysical $\alpha'$, clearly violating
conditions $C_1$, $C_2$, and $C_4$.  We therefore cannot use the $k_{1m}$
solution in Eq. (\ref{k1pm}) but must instead choose the $k_{1p}$ solution.  We
next show that that the discriminant in the
expression for $k_{1p}$ in Eq. (\ref{k1pm}), $D_k = b_2^2-4b_1b_3$, is
nonnegative (actually positive), as it must be. This 
property follows because $b_3 < 0$ in this interval for the representations
under consideration, since $N_{f,b3z} < N_{f,b2z}$ (where we use the relevant
solution of the quadratic equation, labelled $N_{f,b3z,-}$ in Eq. (3.16) of
our \cite{bvh}). Hence, we can write $D_k = b_2^2+4b_1|b_3| > 0$.  
We denote the present scheme transformation with this choice as $S_1$:
\beqs
S_1: \quad\quad & & s_{max} = 1, \quad k_1 = k_{1p} \ , i.e., \cr\cr
                & & a=a'(1+k_{1p}a') \ .
\label{sprime}
\eeqs

Physically, $N_f$ is restricted to take on nonnegative, integral
values. However, since in much of our analysis, we do consider the formal
analytic continuation of $N_f$ from these integral values to positive real
numbers, we remark on one effect of this continuation here.  For a given gauge
group $G$ and fermion representation $R$, if one carries out this analytic
continuation and considers the formal limit $N_f \nearrow N_{f,max}$, i.e., as
one approaches the upper end of the non-Abelian Coulomb phase, as a function of
$N_f$, since $b_1 \to 0$, $k_{1p}$ diverges because of the prefactor
$(2b_1)^{-1}$ in Eq. (\ref{k1pm}).  This divergence in $k_{1p}$ is cancelled in
the actual $S_1$ transformation, which still maps $\alpha_{IR,2\ell} \to 0$ to
$\alpha' \to 0$ as $N_f \nearrow N_{f,max}$. This can be seen by expanding
Eq. (\ref{alfprimesmax1}): $\alpha'_+ \to \sqrt{\alpha/\bar k_{1p}} \to 0$.
Although one does not have to worry about this if one restricts $N_f$ to
physical, integer values in the asymptotically free interval $N_f < N_{f,max}$,
it does lead to significant residual scheme dependence in the comparison
between the four-loop IR zero in the $\overline{MS}$ scheme, and the four-loop
zero computed by applying this $S_1$ scheme transformation to the
$\overline{MS}$ scheme, even for $N_f$ near to $N_{f,max}$.

By construction, since $b'_3=0$ in this scheme, the three-loop zero of
$\beta_{\alpha'}$ is equal to the two-loop zero, as expressed in
Eq. (\ref{s2ttrel}), and as was the case with the $S_2$ and $S_3$ schemes. 
At the four-loop level, the IR zero is given by the physical
(smallest positive) solution of the cubic equation (\ref{cubiceqs2}) with
$b'_4$ given by Eq. (\ref{b4prime}) with $k_1=k_{1p}$ and $k_2=k_3=0$.
We list values of $\alpha'_{IR,n\ell} \equiv \alpha'_{IR,n\ell,S_1}$ in this
$S_1$ scheme, up to $(n=4)$-loop level, as calculated in \cite{scc}, in Table
\ref{betazero_fund_s1s3}, for relevant $N_f$, with fermions in the fundamental
representation and several values of $N$.  For comparison we also include the
values of $\alpha_{IR,n\ell,\overline{MS}}$ for $n=3,4$ in the $\overline{MS}$
scheme from \cite{bvh}.

We carried out the analogous calculations for fermions in the adjoint
representation of SU($N$) in \cite{scc}. Here, $N_{f,b1z}=11/4$ and
$N_{f,b2z}=17/16$, so the only physical, integer value of $N_f \in I$ is
$N_f=2$.  SU(2) models with $N_f=2$ adjoint fermions have been of recent
interest \cite{sanrev}. For both of these cases we found that
\beq
\alpha'_{IR,3\ell,S_1} > \alpha_{IR,3\ell,\overline{MS}}
\label{inequality_3loop_s1}
\eeq
and
\beq
\alpha'_{IR,4\ell,S_1} < \alpha_{IR,4\ell,\overline{MS}}  \ . 
\label{inequality_4loop_s1}
\eeq

For both of these representations, our results obey the required behavior in
Eq. (\ref{alfirprime_upperend}), although one observes that even for rather
large $N_f$ values that are reliably expected to lie in the non-Abelian Coulomb
phase, there is still a significant difference between $\alpha'_{IR,n\ell,S_1}$
and $\alpha_{IR,n\ell,\overline{MS}}$ for $n=3,4$.  We attribute this
difference to the behavior of $k_{1p}$ as a function of $N_f \in I$.  As we
will show, this difference is greater than the corresponding difference when
one uses a scheme transformation such as the $S_{sh_r}$ scheme to be
discussed below.


\section{The $S_{th_r}$ Scheme Transformation}
\label{tanhst}

In this section we study the scheme transformation 
\beq
S_{th_r}: \quad a=\frac{\tanh(ra')}{r} \ . 
\label{tanhgen}
\eeq
Since $\tanh(ra')/r$ is an even function of $r$, we take $r > 0$ with no loss
of generality. The transformation $S_{th_r}$ has the advantage that it depends
on a parameter $r$, which we can vary to study the effect that it has on the
location of the IR fixed point.  In particular, as $r \to 0$, this
transformation smoothly approaches the identity.  The inverse of
Eq. (\ref{tanhgen}) is
\beq
a' = \frac{1}{2r} \, \ln \bigg ( \frac{1+ra}{1-ra} \bigg )
\label{tanhgen_inverse}
\eeq
and the Jacobian is 
\beq
J = \frac{1}{\cosh^2(ra')} \ . 
\label{jtanhgen}
\eeq
In the notation of Eq. (\ref{aap}), 
\beq
f(a') = \frac{\tanh(r a')}{r a'} \ . 
\label{faprime_tanhgen}
\eeq
This has the series expansion of the form (\ref{faprime}), with 
\beq
k_s = 0 \quad {\rm for \ odd} \ \ s 
\label{ksoddzero_tanhgen}
\eeq
and, for even $s$, 
\beq
k_2 = -\frac{r^2}{3} \ , \quad k_4 = \frac{2r^4}{15} \ , 
\label{k24tanhgen}
\eeq
\beq
k_6 = -\frac{17r^6}{315} \ , \quad k_8 = \frac{62r^8}{2835} \ , 
\label{k68tanhgen}
\eeq
and so forth for $k_s$ with higher $s$.

Substituting these expressions for $k_s$ into the general expressions for the
$b'_\ell$, we obtain
\beq
b'_3 = b_3 + \frac{r^2 b_1}{3} \ , 
\label{b3prime_th}
\eeq
\beq
b'_4 = b_4 \ , 
\label{b4prime_th}
\eeq
\beq
b'_5 = b_5 - \frac{r^2 b_3}{3} + \frac{2r^4b_1}{45} \ , 
\label{b5prime_th}
\eeq
\beq
b'_6 = b_6 - \frac{2r^2 b_4}{3} + \frac{r^4 b_2}{15} \ , 
\label{b6prime_th}
\eeq
\beq
b'_7 = b_7 - r^2 b_5 + \frac{r^4 b_3}{5} + \frac{r^6b_1}{315} \ , 
\label{b7prime_th}
\eeq
\beq
b'_8 = b_8 - \frac{4r^2 b_6}{3} + \frac{4r^4 b_4}{9} 
- \frac{4r^6b_2}{189} \ , 
\label{b8prime_th}
\eeq
and so forth for the $b'_\ell$ with $\ell \ge 9$.  .

We apply this $S_{th_r}$ scheme transformation to the $\beta$ function in the
$\overline{MS}$ scheme. We will only need the $b'_\ell$ with $\ell \le 4$ for
this purpose, since (in addition to the scheme-independent $b_1$ and $b_2$) 
only $b_3$ and $b_4$ have been calculated for the $\overline{MS}$ scheme. 
For $N_f$ in the interval $I$ where the two-loop
$\beta$ function has an IR zero, we then calculate the resultant IR zeros in
$\beta_{\alpha'}$ at the three- and four-loop order. We have carried out these
calculations for $N=2,3,4$, with fermions in the fundamental representation and
for a range of $r$ values, namely $r=3, \ 6, \ 9$, and $4\pi \simeq 12.56$.  We
list the results in Table \ref{betazero_fund_tanh}.  For $r=1$, the IR zeros
are almost identical to those in the $\overline{MS}$ scheme and hence are not
listed.  The complex entry for $N=2$, $N_f=7$, $r=4\pi$ is 
$\alpha'_{IR,4\ell,r=4\pi} = 1.718 \pm 0.9285i$.  The presence of this complex
entry is a manifestation of the fact pointed out in \cite{scc} and discussed
further below that this scheme transformation is not acceptable in general. 

As regards the change in the location of the IR zero as a function of the loop
order, we first recall that in \cite{bvh} we showed that for a given $N$ and
$N_f$ (with $N_f \in I$, so the two-loop $\beta$ function has an IR zero), as
one goes from two-loop to three-loop order, the location of this zero decreases
and then as one goes from three-loop to four-loop order, it increases by a
smaller amount, so that the four-loop value is still smaller than the
(scheme-independent) 2-loop value.  Aside from the pathological behavior that
occurs for smaller $N_f$ values where $\alpha_{IR,2\ell}$ gets sufficiently
large (e.g., for $N=2$, $N_f=7$, where $\alpha_{IR,2\ell}=2.83$ and
$\alpha'_{IR,4\ell,r=4\pi}$ is complex), we observe behavior similar to that
which we found in our previous higher-loop calculations for fermions in the
fundamental representation in the $\overline{MS}$ scheme.  First, as is evident
from Table \ref{betazero_fund_tanh}, for a given $N$, $N_f$, and $r$,
\beq
\alpha'_{IR,3\ell,S_{th_r}} < \alpha'_{IR,4\ell,S_{th_r}} < 
\alpha_{IR,2\ell} \quad {\rm for} \ R=fund.
\label{alfir_loop_dep_tanh_fund}
\eeq
These shifts as a function of loop order are
larger for smaller $N_f$ and get smaller as $N_f$ approaches $N_{f,max}$. 
Second, we observe that for a given $N$, $N_f$, and $r$, 
\beq
\alpha'_{IR,n\ell,S_{th_r}} > \alpha_{IR,n\ell,\overline{MS}} \ , 
\quad {\rm for} \ n = 3, \ 4, \quad  R = fund. 
\label{tanhversusmsbar}
\eeq
For a given $N$ and $r$, the values $\alpha'_{IR,n\ell,S_{th_r}}$ approach 
the corresponding $\alpha_{IR,n\ell,\overline{MS}}$ as $N_f \nearrow
N_{f,max}$.  Third, for a given $N$, $N_f$, and loop order
$n=3$ or $n=4$,
\beq
\alpha'_{IR,n\ell,S_{th_r}} \ \  {\rm is \ an \ increasing \ function \ of} \ r
\ .
\label{alfir_tanh_rdependence}
\eeq
For $N_f$ values close to $N_{f,max}$ for a given $N$, these differences in
values are sufficiently small so that the entries may coincide to the given
number of significant figures.

The scheme transformation $S_{th_r}$ with $r=4\pi$ can be written equivalently 
as
\beq
\alpha = \alpha' \tanh(\alpha') \ . 
\label{tanh_alpha}
\eeq
As we pointed out in \cite{scc}, the $S_{th_r}$ scheme transformation with this
value of $r$ is not acceptable, because it violates conditions $C_1$, $C_2$,
and $C_4$.  In particular, as is evident from the inverse of this
transformation, viz.,
\beq
\alpha' = \frac{1}{2} \, \ln \bigg ( \frac{1+\alpha}{1-\alpha} \bigg ) \ ,
\label{tanh_alpha_inverse}
\eeq
the exact inverse transformation maps $\alpha > 1$ to a complex and hence
unphysical, value of $\alpha'$.  At an IR fixed point, it can easily happen
that $\alpha_{IR,2\ell} > 1$, in which case this ST yields a complex,
unphysical $\alpha'$.  For example (see Table III in \cite{bvh}) for $G={\rm
SU}(2)$ with $N_f=8$ fermions in the fundamental representation,
$\alpha_{IR,2\ell}=1.26$ and for SU(3) with $N_f=11$,
$\alpha_{IR,2\ell}=1.23$. More generally, as is evident from
Eq. (\ref{tanhgen_inverse}), the inverse of the scheme transformation
$S_{th_r}$ with a given value of $r$ will map a value $\alpha > 1$ to a
complex, unphysical value of $\alpha'$ if $ r \alpha/(4 \pi) > 1$.  As with the
complex entries in Table \ref{betazero_fund_tanh}, this is another
manifestation of the pathology in this scheme transformation at an IR fixed
point.  In order for this $S_{th_r}$ scheme transformation to satisfy
conditions $C_1$, $C_2$, and $C_4$, it is necessary that for the values of
$\alpha$ of interest,
\beq
r < \frac{4\pi}{\alpha} = \frac{1}{a} \ . 
\label{rcondition_tanh}
\eeq
%


\section{The $S_{sh_r}$ Scheme Transformation}
\label{sinhst} 

In this section we study the scheme transformation 
\beq
S_{sh,r}: \quad a=\frac{\sinh(ra')}{r} \ . 
\label{sinhgen}
\eeq
Since $\sinh(ra')/r$ is an even function of $r$, we take $r > 0$ with no loss
of generality. This has the inverse
\beq
a' = \frac{1}{r} \, \ln \bigg [ ra + \sqrt{1+ (ra)^2} \ \bigg ]
\label{sinhgen_inverse}
\eeq
and the Jacobian
\beq
J = \cosh(ra') \ . 
\label{jsinhgen}
\eeq
In the notation of Eq. (\ref{aap}), 
\beq
f(a') = \frac{\sinh(r a')}{r a'} \ . 
\label{faprime_sinhgen}
\eeq
This has a series expansion of the form (\ref{faprime}) with $k_s=0$ for odd
$s$, as in (\ref{ksoddzero_tanhgen}), and for even $s$,
\beq
k_2 = \frac{r^2}{6} \ , \quad k_4 = \frac{r^4}{120} \ , 
\label{k24sinhgen}
\eeq
\beq
k_6 = \frac{r^6}{5040} \ , \quad k_8 = \frac{r^8}{362880} \ , 
\label{k6sinhgen}
\eeq
and so forth for higher $s$.  

Substituting these expressions for $k_s$ into the general expressions for the
$b'_\ell$, we obtain
\beq
b'_3 = b_3 - \frac{r^2 b_1}{6} \ , 
\label{b3prime_sh}
\eeq
\beq
b'_4 = b_4 \ , 
\label{b4prime_sh}
\eeq
\beq
b'_5 = b_5 + \frac{r^2b_3}{6} + \frac{31r^4b_1}{360} \ , 
\label{b5prime_sh}
\eeq
\beq
b'_6 = b_6 + \frac{r^2 b_4}{3} + \frac{r^4 b_2}{15} \ , 
\label{b6prime_sh}
\eeq
\beq
b'_7 = b_7 + \frac{r^2 b_5}{2} + \frac{3 r^4 b_3}{40} 
- \frac{173 r^6 b_1}{5040} \ , 
\label{b7prime_sh}
\eeq
\beq
b'_8 = b_8 + \frac{2 r^2 b_6}{3} + \frac{r^4 b_4}{9} 
- \frac{4r^6 b_2}{189} \ , 
\label{b8prime_sh}
\eeq
and so forth for the $b'_\ell$ with $\ell \ge 9$.

We apply this $S_{sh_r}$ scheme transformation to the $\beta$ function in the
$\overline{MS}$ scheme. 
For the same reason as was given above, we will only need the $b'_\ell$ with
$\ell \le 4$ for this purpose. For $N_f$ in the interval $I$ where the two-loop
$\beta$ function has an IR zero, we then calculate the resultant IR zeros in
$\beta_{\alpha'}$ at the three- and four-loop order. We have carried out these
calculations for $N=2,3,4$, with fermions in the fundamental representation and
for a range of $r$ values, namely $r=3, \ 6, \ 9$, and $4\pi$. We list the
results in Table \ref{betazero_fund_sinh}.  We denote the IR zero of
$\beta_{\alpha'}$ at the $n$-loop level as $\alpha'_{IR,n\ell}
\equiv \alpha'_{IR,n\ell,S_{sh_r}}$ and in the table we further
shorten this to $\alpha'_{IR,n\ell,r}$. As with the $S_{th_r}$ scheme
transformation, and for the same reason, for $r=1$, the IR zeros are almost
identical to those in the $\overline{MS}$ scheme and hence are not listed.

We observe the following general properties in our calculations of 
$\alpha'_{IR,n\ell,S_{sh_r}}$. First, as is evident
from Table \ref{betazero_fund_sinh}, for a given $N$, $N_f$, and $r$,
\beq
\alpha'_{IR,3\ell,S_{th_r}} < \alpha'_{IR,4\ell,S_{th_r}} < 
\alpha_{IR,2\ell} \quad {\rm for} \ R=fund.
\label{alfir_loop_dep_sinh_fund}
\eeq
As with our calculations with other scheme transformations, these shifts as a
function of loop order are larger for smaller $N_f$ and get smaller as $N_f$
approaches $N_{f,max}$.  Second, for a given $N$, $N_f$, and $r$,
\beq
\alpha'_{IR,n\ell,S_{sh_r}} < \alpha_{IR,n\ell,\overline{MS}} \ , 
\quad {\rm for} \ n = 3, \ 4, \quad  R = fund. 
\label{sinhversusmsbar}
\eeq
For a given $N$ and $r$, the values $\alpha'_{IR,n\ell,S_{sh_r}}$ approach 
the corresponding $\alpha_{IR,n\ell,\overline{MS}}$ as $N_f \nearrow
N_{f,max}$.  Third, for a given $N$, $N_f$, and loop order $n=3$ or $n=4$,
\beq
\alpha'_{IR,n\ell,S_{th_r}} \ \  {\rm is \ a \ decreasing \ function \ of} \ r
\ .
\label{alfir_sinh_rdependence}
\eeq
Note that the inequalities (\ref{sinhversusmsbar}) and 
(\ref{alfir_sinh_rdependence}) are opposite to (\ref{tanhversusmsbar}) and
(\ref{alfir_tanh_rdependence}) for the $S_{th_r}$ scheme transformation. 
As was the case with the other schemes, for $N_f$ values close to $N_{f,max}$
for a given $N$, these properties are sufficiently small so that the entries
may coincide to the given number of significant figures.

In contrast with the $S_{th_r}$ scheme transformation, the $S_{sh_r}$
transformation is acceptable for $r$ values up to the largest that we consider,
viz., $r=4\pi$, where it takes the form
\beq
\alpha = \sinh(\alpha') \ .
\label{alphasinh}
\eeq
This is understandable since the inverse transformation, 
(\ref{sinhgen_inverse}), is not singular, whereas the inverse of the 
$S_{th_r}$ transformation, (\ref{sinhgen_inverse}), is singular for 
for $\alpha \to 1$ for this value of $r$.  As with the other scheme
transformations, the three- and four-loop values of the IR zero in the 
$S_{sh_r}$ scheme approach the corresponding values in the $\overline{MS}$ 
as $N_f \to N_{f,max}$, in accord with Eq. (\ref{alfirprime_upperend}). 

Some comparative remarks are in order concerning the $S_1$ and $S_{sh_r}$ 
scheme transformations.  We find that the $S_{sh_r}$ scheme transformation with
moderate $r$ leads to smaller shifts in the location of the IR zero than was
the case with the $S_1$ scheme transformation, when both are applied to 
the $\beta$ function in the $\overline{MS}$ scheme.  We have explained the
origin of this as resulting from a particular feature of the parameter 
$k_{1p}$ that enters in the $S_1$ scheme transformation. In general, we find
that even for smaller $N_f$ values (lying above $N_{f,b2z}$) the $S_{sh_r}$
transformation with moderate $r$ produces rather small shifts in the location
of the IR zero.  For example (cf. Table \ref{betazero_fund_sinh}), 
for SU(3) with $N_f=10$, we obtain the following fractional shifts in this IR
zero at the three-loop and four-loop level: 
\beqs
& & \frac{\alpha'_{IR,3\ell,S_{sh_r},r=4\pi} - \alpha_{IR,3\ell,\overline{MS}}}
{\alpha_{IR,3\ell,\overline{MS}}} = -0.054  \cr\cr
& & \cr\cr
& & 
\frac{\alpha'_{IR,4\ell,S_{sh_r},r=4\pi} - \alpha_{IR,4\ell,\overline{MS}}}
{\alpha_{IR,4\ell,\overline{MS}}} = -0.065 \cr\cr
& & {\rm for} \ {\rm SU}(3), \ N_f=10, \ r=4\pi \ . 
\label{alfshiftsu3nf10sinh34loop}
\eeqs
One would thus tend to prefer the $S_{sh_r}$ scheme transformation, since it
minimizes scheme dependence at higher-loop order.  However, the $S_1$
transformation provides an example of how there may still be significant
dependence when one uses certain scheme transformations.  We will show another
example of this in the next section, using an illustrative exact $\beta$
function, for which a slight change in $r$ in the $S_{sh_r}$ transformation can
have a significant effect on the nature of an IR zero at the three-loop order.


\section{Study with An Illustrative Exact $\beta$ Function}
\label{exactbeta}

It is instructive to study series expansions of an illustrative hypothetical
exact $\beta$ function in order to ascertain the accuracy and reliability of
finite-order analyses and the effects of scheme transformations.  Here we shall
take one such function, which has an exactly known infrared zero that is
reached from the origin.  It should be emphasized at the outset that, although
the function that we use in Eq. (\ref{betah}) with (\ref{sinxoverx}) below is
designed to emulate some properties of the $\beta$ function of an
asymptotically free non-Abelian gauge theory with fermions, we do not mean to
imply that it is fully realistic.  Instead, we use it in the spirit of a
reasonable test function which embodies some relevant features and can serve as
a theoretical laboratory in which to investigate how well analyses of truncated
series expansions probe the IR zero and how this is affected by scheme
transformations. 

Because we are interested in the evolution of an asymptotically free theory
from the neighborhood of the UV fixed point at $\alpha=0$ to an IR fixed point,
we require that this illustrative $\beta$ function have the property that, as
$\alpha$ increases from zero, it has a zero at a finite value of $\alpha$,
which we denote as $\alpha_{IR}$.  We also require that it be bounded in the
interval
\beq
0 \le \alpha \le \alpha_{IR} \ . 
\label{alphainterval}
\eeq
It is convenient to define a scaled quantity
\beq
\tilde \alpha \equiv \frac{\alpha}{\alpha_{IR}} \ . 
\label{tildealpha}
\eeq
Since we assume that the evolution of the theory from the UV to the IR starts
from a small value in the UV, we only need to consider the behavior of $\beta$
in this interval (\ref{alphainterval}). From Eq. (\ref{beta}), the $\beta$
function has the form, for small $\alpha$ in the deep UV,
\beqs
\beta_\alpha & = & -2\bar b_1\alpha^2 \bigg [ 1+ \frac{\bar b_2}{\bar b_1} \,
 \alpha + O(\alpha^2) \bigg ] \cr\cr
    & = & -2\bar b_1 \alpha^2 \bigg [ 1 - \frac{\alpha}{\alpha_{IR,2\ell}} + 
O(\alpha^2) \bigg ] \ . 
\label{eqs2loop}
\eeqs 
In general, we can write
\beq
\beta_{\alpha} = -2 \bar b_1 \alpha^2 h(\alpha) \ , 
\label{betah}
\eeq
where the function $h(\alpha)$ satisfies
\beq
h(0)=1 \ . 
\label{hzero1}
\eeq

{\it A priori}, one could consider functions $h(\alpha)$ with either a finite
or an infinite series expansion. We shall consider an illustrative example of
the latter case, namely
\beq
h(\alpha) = \frac{\sin ( \pi \sqrt{\tilde \alpha} \ ) }
    {( \pi \sqrt{\tilde \alpha} \ ) } \ . 
\label{sinxoverx}
\eeq
Here we use $\sqrt{\tilde \alpha }$ because $\sin(x)/x$ has only even
powers in its Taylor series expansion
\beq
\frac{\sin x}{x} = \sum_{n=0}^\infty (-1)^n \frac{x^{2n}}{(2n+1)!} \ , 
\label{sinxoverxseries}
\eeq
but we want a $\beta$ function with odd, as well as even, powers of $\alpha$,
to emulate a typical $\beta$ function encountered in a non-Abelian gauge
theory.  (One could equally well use a similar trigonometric function with this
property (and the property (\ref{hzero1})), such as $h(\alpha) =
\cos[(\pi/2)\sqrt{\tilde \alpha} \ ]$).  As noted above, the feature that
(\ref{sinxoverx}) and this cosine function have an infinite number of zeros
beyond the one at $\tilde \alpha = 1$, i.e., $\alpha=\alpha_{IR}$, will not be
of direct concern to us, since we are only interested in their behavior in the
interval (\ref{alphainterval}).  Although the illustrative $\beta$ function in
Eq. (\ref{betah}) with (\ref{sinxoverx}) has no explicit $N_f$-dependence, one
may regard it as implicitly incorporating this through the value of
$\alpha_{IR}$. 

Substituting (\ref{sinxoverxseries}) into (\ref{betah}), we have, for this
illustrative $\beta$ function,
\beq
\beta_{\alpha} = -2\bar b_1 \alpha^2 \, \sum_{\ell=1}^\infty 
\frac{(-\pi^2\alpha/\alpha_{IR})^{\ell-1}}{(2\ell-1)!} \ . 
\label{betasinseries}
\eeq
Hence, in the notation of Eq. (\ref{beta}), 
\beq
\frac{\bar b_\ell}{\bar b_1} = \frac{(-\pi^2/\alpha_{IR})^{\ell-1}}{(2\ell-1)!}
\label{bbarellsin}
\eeq
or equivalently, 
\beq
\frac{b_\ell}{b_1} = \frac{(-4\pi^3/\alpha_{IR})^{\ell-1}}{(2\ell-1)!} \ . 
\label{bellsin}
\eeq

Before performing a scheme transformation, we first analyze finite-order
truncations of this $\beta$ function to see how closely the resulting
determination of the IR zero compares with the exact value, $\alpha_{IR}$.
Obviously, no claim is made that this $\beta$ function actually arose from a
loop calculation, but it will be useful to employ the terminology of loops to
refer to the expansion order.  To four-loop order, $\ell=4$, Eq. (\ref{beta})
reads 
\beq
\beta_{\alpha}  =  -2\bar b_1 \alpha^2 \, \Big [ 1 +
\frac{\bar b_2}{\bar b_1}\, \alpha + 
\frac{\bar b_3}{\bar b_1}\, \alpha^2 +
\frac{\bar b_4}{\bar b_1}\, \alpha^3 + O(\alpha^4) \Big ] \ . 
\label{betasinfourloop}
\eeq
Explicitly, 
\beq
\beta_{\alpha} = -2\bar b_1 \tilde\alpha^2 \, 
\Big [ 1- \frac{\pi^2}{3!} \, \tilde\alpha + 
\frac{\pi^4}{5!} \, \tilde\alpha^2 - \frac{\pi^6}{7!} \, \tilde\alpha^3 + 
O(\tilde\alpha^4) 
\Big ] \ . 
\label{betasinexplicit}
\eeq
For our further discussion, we shall define a compact notation consistent with 
Eq. (\ref{tildealpha}), namely 
\beq
\tilde \alpha_{IR,n\ell} \equiv \frac{\alpha_{IR,n\ell}}{\alpha_{IR}} \ .
\label{tildealf_nloop}
\eeq
At the two-loop order, the $\beta$ function given in Eqs. 
(\ref{betasinfourloop}) and (\ref{betasinexplicit})  has an IR zero at
$\alpha_{IR,2\ell} = -\bar b_1 /\bar b_2 = (6/\pi^2)\, \alpha_{IR} = 
0.60793 \, \alpha_{IR}$, i.e., 
\beq
\tilde\alpha_{IR,2\ell} = 0.60793 \ , 
\label{alfir_2loop_sin}
\eeq
to the indicated numerical accuracy. Evidently, this two-loop estimate of the
IR zero differs substantially from the exact value of the IR zero, being
approximately 40 \% smaller than this value.  Interestingly, at the three-loop
level, although $\beta_\alpha$ has two zeros at nonzero values of
$\tilde\alpha$, neither of them is a physical IR zero; instead, they form the
complex-conjugate pair
\beq
\tilde\alpha_{IR,3\ell,\pm} = \frac{2(5 \pm \sqrt{5} \, i)}{\pi^2} = 
 1.0132 \pm 0.4531i \ . 
\label{alfir_3loop_sin}
\eeq
This is an important result, since it illustrates the basic fact from calculus
that a polynomial obtained as a truncation of a series expansion for a given
function does not necessarily accurately reproduce the zeros of that function.
In the present case, the real part of the complex pair of zeros is 
rather close to 1, but the imaginary part is not small relative to
this real part, so that in the complex plane, the distance of each of these
roots from 1, i.e., the distance of the roots in $\alpha$ from $\alpha_{IR}$,
is substantial.  At four-loop order, the
$\beta_\alpha$ function has three nonzero roots in $\tilde\alpha$, namely a 
physical IR zero close to the exact value,
\beq
\tilde\alpha_{IR,4\ell} = 0.9603 \ , 
\label{alfir4loop_sin}
\eeq
about 4 \% smaller than the exact value, together with a complex conjugate pair
at $\tilde \alpha = 1.6476 \pm 1.6566i$. 

We have continued this analysis up to $(n=8)$-loop order.  At five-loop
level, the equation $\beta_\alpha=0$ has a real root very close to the exact
value,
\beq
\tilde\alpha_{IR,5\ell} = 1.0045 \ , 
\label{alfir_5loop_sin}
\eeq
together with another real root at $\tilde\alpha=2.4958$ and a complex
pair, $\tilde\alpha=1.8974 \pm 3.4138i$.  At the six-loop level, the
equation $\beta_\alpha=0$ has five nonzero solutions for $\tilde\alpha$, namely
\beq
\tilde\alpha_{IR,6\ell} = 0.99972 \ , 
\label{alfir_6loop_sin}
\eeq
and two pairs of complex-conjugate roots.  At the seven-loop level,
the equation $\beta_\alpha=0$ has the root 
\beq
\tilde\alpha_{IR,7\ell} = 1.00001346 \ , 
\label{alfir_7loop_sin}
\eeq
together with two pairs of complex-conjugate roots and a larger positive real
root at $\tilde\alpha=3.621288$. Finally, at the eight-loop level, the
equation $\beta_\alpha=0$ yields 
\beq
\tilde\alpha_{IR,8\ell} = 0.999999507 \ , 
\label{alfir_8loop_sin}
\eeq
together with two pairs of complex-conjugate roots and two larger real roots. 
These values of the physical IR zero for $4 \le \ell \le 8$ yield the following
fractional differences with respect to the exact value:
\beq
\tilde\alpha_{IR,4\ell}-1 \equiv 
\frac{\alpha_{IR,4\ell}-\alpha_{IR}}{\alpha_{IR}} = -3.97 \times 10^{-2} \ , 
\label{dif4loopsin}
\eeq
\beq
\frac{\alpha_{IR,5\ell}-\alpha_{IR}}{\alpha_{IR}} = 4.52 \times 10^{-3} \ ,
\label{dif5loopsin}
\eeq
\beq
\frac{\alpha_{IR,6\ell}-\alpha_{IR}}{\alpha_{IR}} = -2.83 \times 10^{-4}  \ ,
\label{dif6loopsin}
\eeq
\beq
\frac{\alpha_{IR,7\ell}-\alpha_{IR}}{\alpha_{IR}} = 1.35 \times 10^{-5} \ ,
\label{dif7loopsin}
\eeq
and
\beq
\frac{\alpha_{IR,8\ell}-\alpha_{IR}}{\alpha_{IR}} = -0.493 \times 10^{-6})  \ .
\label{dif8loopsin}
\eeq
Thus, once one gets beyond the three-loop order, these values converge
monotonically toward the exact value of $\alpha_{IR}$. 

We next perform a scheme transformation on $\beta_\alpha$ and study the shift
in the values of the IR zero of $\beta_{\alpha'}$, calculated to the various
orders considered here.  We denote these as $\alpha'_{IR,n\ell}$ and the ratios
with respect to $\alpha_{IR}$ as $\tilde\alpha'_{IR,n\ell}$. 
For definiteness, we use the $S_{sh_r}$ transformation
given in Eq. (\ref{sinhgen}), i.e., $\alpha = (4\pi/r)\sinh(r\alpha'/(4\pi))$, 
with variable $r$.  As noted before, without loss of generality, we may take $r
> 0$. Clearly, as $r \to 0^+$, the $S_{sh_r}$ scheme transformation approaches
the identity map, so, by continuity, in this limit, the resulting values of the
IR zero calculated at the $\ell$-loop level approach those obtained
above. However, as we will show next, the values that one gets for larger $r$
depend sensitively on this parameter. Of course, at the two-loop level, 
since $b'_\ell=b_\ell$ for $\ell=1,2$, we get the zero at the same place, but 
now in the $\alpha'$ variable, namely, 
\beq
\tilde\alpha'_{IR,2\ell,S_{sh_r}} = 0.60793 \ . 
\label{tildealf_2loop_sin}
\eeq

At the three-loop level, the condition $\beta_{\alpha'}=0$ yields (aside from
the double root at $\alpha'=0$ corresponding to the UV fixed point), the 
quadratic equation
\beq
1 - \frac{\pi^2}{6} \, \tilde\alpha' + 
\Big ( \frac{\pi^4}{120}-\frac{r^2}{96\pi^2} \Big ) \, (\tilde\alpha')^2=0 \ . 
\label{eq3primeloop}
\eeq
This equation obviously has a singular behavior at the value of $r$ that causes
the coefficient of the $(\tilde\alpha')^2$ term to vanish, namely $r =
2\pi^3/\sqrt{5} = 27.73..$.  We assume that $r$ does not take on this
value. The equation then has the two formal solutions,
\beq
\frac{\tilde\alpha'_{IR,3\ell}}{4 \pi} = 
\frac{20\pi^3 \pm \sqrt{150r^2-20\pi^6}} {2(4\pi^6-5r^2)} \ . 
\label{eq3looprootssinh}
\eeq
We showed above that in the analysis of $\beta_\alpha$ at this three-loop
level, there are no real roots.  Here, in contrast, for sufficiently large $r$,
these roots become real.  This demonstrates how a scheme transformation can
qualitatively, as well as quantitatively, change the analysis of the IR zero of
a $\beta$ function.  In the present case, the roots are real if the
discriminant is nonnegative, i.e, if
\beq
r \ge \Big ( \frac{2}{15} \Big )^{1/2} \, \pi^3 = 11.322 \ . 
\label{rineq}
\eeq
In order to get real roots at the three-loop level, we restrict to $r$ values
that satisfy this inequality.  For example, let us take $r=4\pi$.  Then from
Eq. (\ref{eq3looprootssinh}) we obtain an IR zero at 
\beq
\tilde\alpha'_{IR,3\ell,S_{sh_r}} = 1.000400 \quad {\rm for} \ \ r=4\pi
\label{alfirprime_3loop_sin}
\eeq
together with another real root $\tilde\alpha' = 1.54959$.  Although the
three-loop value in Eq. (\ref{eq3looprootssinh}) is very close to the exact
value 1, i.e., $\alpha'_{IR,3\ell,S_{sh_r}}$ is very close to $\alpha_{IR}$, 
this is fortuitous.  For example, if one increases $r$
from $4\pi = 12.566$ slightly to $r=15$, the value in
Eq. (\ref{alfirprime_3loop_sin}) shifts to $\tilde\alpha_{IR,3\ell,S_{sh_r}} =
1.19414$.  If, on the other hand, one decreases $r$ to ostensibly reasonable
values below the lower bound (\ref{rineq}), then one would revert back to the
situation encountered in the analysis of $\beta_\alpha$, namely there would not
be any physical IR zero at this three-loop level.

 At the four-loop level, if one continues to use the value $r=4\pi$, the
condition $\beta_{\alpha'}=0$ yields one real root, which is the IR zero,
\beq
\tilde\alpha'_{IR,4\ell,S_{sh_r}} = 0.79922 \quad {\rm for} \ \ r=4\pi \ , 
\label{alfirprime_4loop_sin}
\eeq
together with a pair of complex-conjugate roots.  One can carry this analysis
to higher-loop level.  For example, at five-loop level, with $r=4\pi$, the
condition $\beta_{\alpha'}=0$ yields not real solutions for an IR zero, but 
instead a quartic equation with two pairs of complex-conjugate roots.  

In closing this section, we again emphasize that we have carried out this
analysis in the spirit of using a test function with reasonable behavior in the
relevant interval (\ref{alphainterval}) to study how well analyses of a finite
series expansion probe its IR zero, and the effect of a scheme transformation
on these.  There is obviously no implication that other properties of the  
particular test function (\ref{betah}) with (\ref{sinxoverx}) (such 
as the infinitely many zeros at $\sqrt{\tilde \alpha} = s$ with $s \ge 2$) 
are relevant to the true $\beta$ function of a non-Abelian gauge theory. 


\section{Anomalous Dimension of Fermion Bilinear}
\label{gamma}

The anomalous dimension $\gamma_m$ describes the scaling of a fermion bilinear
and the running of a dynamically generated fermion mass in the phase with
spontaneous chiral symmetry breaking. It plays an important role in technicolor
theories, via the renormalization group factor $\eta = \exp[\int dt \,
\gamma_m(\alpha(t))]$ that can enhance dynamically generated Standard-Model
fermion masses.  In the non-Abelian Coulomb phase (which is a conformal phase),
the IR zero of $\beta$ is exact, although a calculation of it to a finite-order
in perturbation theory is only approximate, and $\gamma_m$ evaluated at this IR
fixed point is exact. In the phase with S$\chi$SB, where an IR fixed point, if
it exists, is only approximate, $\gamma_m$ is an effective quantity describing
the running of a dynamically generated fermion mass for the evolution of the
theory near this approximate IRFP.  In \cite{bvh} we evaluated $\gamma_m$ to
three- and four-loop order at the IR zero of $\beta$ calculated to the same
order and showed that these higher-loop results were somewhat smaller than the
two-loop evaluation.  In both the conformal and nonconformal phases it is
important to assess the scheme-dependence of $\gamma_m$ when calculated to
finite order.  At an exact zero of $\beta$, the anomalous dimension
$\gamma_m(\alpha)$ calculated in a given scheme is the same as the anomalous
dimension $\gamma_m'(\alpha')$ calculated in another scheme \cite{gross75}.
Our results in \cite{scc} and here concerning shifts in the location of the IR
zero resulting from a scheme transformation show that, {\it a priori}, a
transformation may introduce significant shifts in both this location and in
the resultant value of $\gamma_m$, especially when the IR fixed point occurs at
moderate to strong coupling.  For a given gauge group $G$ and fermion
representation $R$, the value of $\alpha_{IR,2\ell}$ gets larger as $N_f
\searrow N_{f,b2z}$, and hence, understandably, the shift in
$\alpha'_{IR,n\ell}$ when calculated in a different scheme can be
significant. The same comment applies to $\gamma_m$, although part of this
region of $N_f \gsim N_{f,b2z}$ is in the phase with spontaneous chiral
symmetry breaking rather than the chirally symmetric phase, so the IR fixed
point is only approximate.  For a well-behaved scheme transformation such as
$S_{sh_r}$ with moderate $r$, as $N_f$ increases throughout the non-Abelian
Coulomb phase, this scheme-dependent shift in a finite-loop-order calculation
of the IR zero and resultant shift in the value of $\gamma_m$, calculated to
the same finite-loop order, become small.


\section{Discussion and Conclusions}
\label{discussion}

In this paper, extending the work in \cite{scc}, we have given a detailed
analysis of the effects of scheme transformations in the vicinity of an exact
or approximate infrared fixed point in an asymptotically free gauge theory with
fermions.  We have discussed a set of necessary conditions that such
transformations must obey and have shown with several examples that, although
these can easily be satisfied in the vicinity of an ultraviolet fixed point,
they constitute significant restrictions on scheme transformations at an
infrared fixed point.  This is especially true when this fixed point occurs at
a relatively strong coupling. 

We have constructed acceptable scheme transformations and have used these to
study the scheme-dependence of an infrared fixed point, making comparison with
our previous three-loop and four-loop calculations of the location of this
point in the $\overline{MS}$ scheme in \cite{bvh}.  The $S_1$ transformation,
which renders the three-loop coefficient of the $\beta_{\alpha'}$ function
zero, provides an example of how a scheme transformation can produce
significant scheme dependence in an IR zero.  The $S_{sh_r}$ scheme
transformation with moderate $r$ is better behaved than the $S_1$
transformation and introduces smaller scheme-dependent shifts in the location
of the IR zero. This $S_{sh_r}$ transformation with moderate $r$ provides a
valuable tool to assess scheme dependence. As applied to the $\beta$ function
in the $\overline{MS}$ scheme, it shows that this dependence is small in the
vicinity of both the UV fixed point at $\alpha=0$ and an IR fixed point at
sufficiently small coupling.  It also gives a quantitative measure of the size
of the scheme-dependence in the calculation of this fixed point at the
three-loop and four-loop order, both at small and at larger couplings.

We have constructed an illustrative exact $\beta$ function of an asymptotically
free theory with an infrared zero and have used it as a theoretical laboratory
in which to assess the accuracy with which finite-order truncations of the
series expansion of this $\beta$ function are able to determine the IR zero.
Applying the $S_{sh_r}$ scheme transformation to the series expansion for this
illustrative $\beta$ function, we have also studied the consequences of this
for the determination of the IR zero in the $\alpha'$ variable from a
finite-order truncation of the series.  For the illustrative $\beta$ function,
we find that this scheme transformation can have a significant effect, 
especially at low orders in the expansion. 

We believe that the results reported here give a deeper insight into scheme
transformations of the $\beta$ function and scheme-dependence of infrared fixed
points in non-Abelian gauge theories with fermions.  There is clearly more
interesting work to be done investigating this question.  The knowledge gained
will be useful for a better understanding of the UV to IR evolution of these
theories, in particular, those with fermion contents that result in
quasi-conformal behavior.

\begin{acknowledgments}

This research was partially supported by a Sapere Aude Grant (T.A.R.) and
NSF grant NSF-PHY-09-69739 (R.S.).

\end{acknowledgments}


\section{Appendix}

In this appendix we first give the expressions that we have calculated for 
$b'_\ell$ with $\ell=6,7,8$: 
\begin{widetext}
\beqs
b_6' & = & b_6 +4k_1b_5+(4k_1^2+2k_2)b_4+4k_1k_2b_3 
 + (2k_1^4-6k_1^2k_2+4k_1k_3+3k_2^2-2k_4)b_2 \cr\cr
     & + & (-8k_1^5+28k_1^3k_2-16k_1^2k_3-20k_1k_2^2 
      + 8k_1k_4+12k_2k_3 -4k_5)b_1 \ ,
\label{b6prime}
\eeqs
\beqs
b_7' & = & b_7 +5k_1b_6+(7k_1^2+3k_2)b_5+
     (2k_1^3+7k_1k_2+k_3)b_4 + (k_1^4-2k_1^2k_2+4k_1k_3+3k_2^2-k_4)b_3 \cr\cr
     & + & (-4k_1^5+15k_1^3k_2-9k_1^2k_3-12k_1k_2^2+9k_2k_3
+5k_1k_4-3k_5)b_2 \cr\cr
     & + & (16k_1^6-68k_1^4k_2+40k_1^3k_3-21k_1^2k_4
+73k_1^2k_2^2-58k_1k_2k_3+10k_1k_5+16k_2k_4-12k_2^3+9k_3^2-5k_6)b_1 \ , 
\cr\cr
& &
\label{b7prime}
\eeqs
and
\beqs
b_8' & = & b_8 +6k_1b_7+(11k_1^2+4k_2)b_6+
     (6k_1^3+12k_1k_2+2k_3)b_5 + (k_1^4+4k_1^2k_2+6k_1k_3+4k_2^2)b_4 \cr\cr
     & + & (-2k_1^5+8k_1^3k_2-4k_1^2k_3-6k_1k_2^2+8k_2k_3
+4k_1k_4-2k_5)b_3 \cr\cr
     & + & (8k_1^6-36k_1^4k_2+22k_1^3k_3-12k_1^2k_4
+42k_1^2k_2^2-36k_1k_2k_3+6k_1k_5+12k_2k_4-8k_2^3+7k_3^2-4k_6)b_2 \cr\cr
   & + & (-32k_1^7+160k_1^5k_2-96k_1^4k_3+52k_1^3k_4-230k_1^3k_2^2
   - 26k_1^2k_5 + 208k_1^2k_2k_3 \cr\cr
  & + & 12k_1k_6+84k_1k_2^3-42k_1k_3^2
-76k_1k_2k_4 + 20k_2k_5+24k_3k_4-52k_2^2k_3-6k_7)b_1 \ . 
\cr\cr
& &
\label{b8prime}
\eeqs

\end{widetext}

For reference, we list the expressions for $b_1$ \cite{b1},  $b_2$
\cite{b2}, and, in the $\overline{MS}$ scheme, $b_3$ \cite{b3}, calculated for
a vectorial gauge theory with $N_f$ (massless) fermions transforming according
to the representation $R$ of the gauge group $G$ \cite{casimir}:
\beq
b_1 = \frac{1}{3}(11 C_A - 4T_fN_f)
\label{b1}
\eeq
\beq
b_2=\frac{1}{3}\left [ 34 C_A^2 - 4(5C_A+3C_f)T_f N_f \right ]
\ .
\label{b2}
\eeq
\begin{widetext}
\beqs
& & b_3 = \frac{2857}{54}C_A^3 +
+ T_f N_f \bigg[ 2C_f^2 - \frac{205}{9} C_AC_f - \frac{1415}{27}C_A^2 \bigg ]
 + (T_f N_f)^2 \bigg [ \frac{44}{9}C_f + \frac{158}{27}C_A \bigg ] \ .
\cr\cr
& &
\label{b3}
\eeqs
In our calculations we have also used the $\overline{MS}$ result for $b_4$
\cite{b4}, but we do not list it here because of its length.
\end{widetext}

The interval $I$ in which the two-loop $\beta$ function has an IR zero is given
in Eq. (\ref{nfinterval}).  The lower end of this interval is defined by the
condition that $b_2$ decreases through zero, which occurs at the value
$N_f=N_{f,b2z}$ given in Eq. (\ref{nfb2z}).  Numerical values of $\bar b_\ell$
were presented in \cite{bvh}, e.g., for the fundamental representation in Table
I of that reference. As discussed in \cite{bvh}, for $N_f=0$ and sufficiently
small, $b_2$, $b_3$, and $b_4$ are all positive, and they decrease with
increasing $N_f$.  The value of $N_f$ at which $b_3$ goes through zero and
becomes negative, denoted $N_{f,b3z}$, is smaller than the value $N_{f,b2z}$,
so that $b_3$ is generically negative in the interval $I$
(cf. Eq. (\ref{nfinterval})) where the two-loop $\beta$ function has an IR
zero.  As is evident in Table I, the four-loop coefficient $b_4$ can be
positive or negative in this interval $I$. The upper end of the interval $I$
occurs at $N_f = N_{f,b1z} = N_{f,max}$ \cite{nfreal}, where $b_1 \to
0^+$. The values of $b_2$ and $b_3$ at $N_f=N_{f,max}$ are used implicitly in
the text, in particular, in our discussion of the $S_1$ scheme transformation,
so we list them here:
\beq
b_2 = -C_A(7C_A+11C_f) \quad {\rm at} \ \ N_f = N_{f,max}
\label{b2nfmax}
\eeq
and
\beq
b_3 = -\frac{C_A}{24}(1127C_A^2+616C_AC_f-132C_f^2) \ \  {\rm at} \ \ 
N_f = N_{f,max}
\label{b3nfmax}
\eeq
We denote these as $(b_2)_{N_{f,max}}$ and $(b_3)_{N_{f,max}}$, respectively. 
For the fundamental representation, 
\beq
(b_2)_{N_{f,max},fund.} = - \bigg [ \frac{(5N)^2-11}{2} \bigg ] 
\label{b2nfmax_fund}
\eeq
and
\beq
(b_3)_{N_{f,max},fund..} = - \bigg [ \frac{1402N^4-242N^2-33}{24N} \bigg ] \ .
\label{b3nfmax_fund}
\eeq
These are both negative for all physically relevant $N$. Specifically, 
with $N_f$ continued from the nonnegative integers to the nonnegative reals, 
\beq
(b_2)_{N_{f,max},fund.} < 0 \quad {\rm for} \ \ 
N > \frac{\sqrt{11}}{5} = 0.66332..
\label{b2nfmaxfundgt}
\eeq
and 
\beqs
& & (b_3)_{N_{f,max},fund.} < 0 \cr\cr
& & {\rm for} \ \ 
N > \frac{\Big [ 169642 + 9814\sqrt{1243} \ \Big ]^{1/2}}{1402} = 0.512186..
\cr\cr
& & 
\label{b3nfmaxfundgt}
\eeqs
For the adjoint representation, 
\beq
(b_2)_{N_{f,max}, adj.} = -18N^2 
\label{b2nfmax_adj}
\eeq
and
\beq
(b_3)_{N_{f,max}, fund.} = - \frac{537N^3}{8} \ , 
\label{b3nfmax_adj}
\eeq
which are also negative. 

This research was partially supported by a Sapere Aude Grant (TAR) and 
NSF grant NSF-PHY-09-69739 (RS).



\newpage

\begin{table}
\caption{\footnotesize{Values of the IR zeros of $\beta_{\alpha}$ in the
$\overline{MS}$ scheme and the respective $\beta_{\alpha'}$ functions obtained
by applying the $S_1$, $S_2$, and $S_3$ scheme transformations to the
$\overline{MS}$ $\beta_{\alpha}$ function.  The listings are for an ${\rm
SU}(N)$ gauge theory with $N_f$ (massless) fermions in the fundamental
representation, for $N=2,3,4$, calculated to $n$-loop order and denoted as
$\alpha_{IR,n\ell,\overline{MS}}$ and $\alpha'_{IR,n\ell,S_i}$, where $i=1, \ 
2, \ 3$, respectively.  Here,
$\alpha_{IR,2\ell,\overline{MS}}=\alpha'_{IR,2\ell,S_i}$ is scheme-independent,
so we denote it simply as $\alpha_{IR,2\ell}$. Since all of these $S_i$ scheme
transformations with $i=1, \ 2, \ 3$ yield $b'_3=0$, it follows that
$\alpha'_{IR,3\ell,S_i}=\alpha'_{IR,2\ell}=\alpha_{IR,2\ell}$. The notation
n.p. means not physical, i.e., there is no physical solution for 
$\alpha'_{IR,4\ell,S_i}$. See text for further details.}}
\begin{center}
\begin{tabular}{|c|c|c|c|c|c|c|c|} \hline\hline
$N$ & $N_f$ & $\alpha_{IR,2\ell}$ & $\alpha_{IR,3\ell,\overline{MS}}$ 
& $\alpha_{IR,4\ell,\overline{MS}}$ 
& $\alpha'_{IR,4\ell,S_1}$ & $\alpha'_{IR,4\ell,S_2}$ & 
$\alpha'_{IR,4\ell,S_3}$ \\
\hline
 2  &  7  &  2.83   & 1.05   & 1.21   & 0.640  & n.p.   &  0.488  \\
 2  &  8  &  1.26   & 0.688  & 0.760  & 0.405  & n.p.   &  0.633  \\
 2  &  9  &  0.595  & 0.418  & 0.444  & 0.2385 & n.p.   &  0.730  \\
 2  & 10  &  0.231  & 0.196  & 0.200  & 0.109  & 0.240  &  0.248  \\
 \hline
 3  & 10  &  2.21   & 0.764  & 0.815  & 0.463  & n.p.   &  0.316  \\
 3  & 11  &  1.23   & 0.578  & 0.626  & 0.344  & n.p.   &  0.391  \\
 3  & 12  &  0.754  & 0.435  & 0.470  & 0.254  & n.p.   &  0.444  \\
 3  & 13  &  0.468  & 0.317  & 0.337  & 0.181  & n.p.   &  0.4385 \\
 3  & 14  &  0.278  & 0.215  & 0.224  & 0.121  & 0.321  &  0.358  \\
 3  & 15  &  0.143  & 0.123  & 0.126  & 0.068  & 0.148  &  0.152  \\
 3  & 16  &  0.042  & 0.040  & 0.040  & 0.0215 & 0.042  &  0.042  \\
\hline
 4  & 13  &  1.85   & 0.604  & 0.628  & 0.365  & n.p.   &  0.228  \\
 4  & 14  &  1.16   & 0.489  & 0.521  & 0.293  & n.p.   &  0.276  \\
 4  & 15  &  0.783  & 0.397  & 0.428  & 0.235  & n.p.   &  0.311  \\
 4  & 16  &  0.546  & 0.320  & 0.345  & 0.187  & n.p.   &  0.339  \\
 4  & 17  &  0.384  & 0.254  & 0.271  & 0.146  & n.p.   &  0.362  \\
 4  & 18  &  0.266  & 0.194  & 0.205  & 0.110  & n.p.   &  n.p.   \\
 4  & 19  &  0.175  & 0.140  & 0.145  & 0.0785 & 0.193  &  0.208  \\
 4  & 20  &  0.105  & 0.091  & 0.092  & 0.050  & 0.108  &  0.111  \\
 4  & 21  &  0.047  & 0.044  & 0.044  & 0.023  & 0.048  &  0.048  \\
\hline\hline
\end{tabular}
\end{center}
\label{betazero_fund_s1s3}
\end{table}


\begin{widetext}

\begin{table}
\caption{\footnotesize{Values of the IR zeros of $\beta_{\alpha}$ in the
$\overline{MS}$ scheme and $\beta_{\alpha'}$ after applying the $S_{th_r}$
scheme transformation to the $\overline{MS}$ scheme, for an
SU($N$) theory with $N_f$ fermions in the fundamental representation, for
$N=2,3,4$, calculated to $n$-loop order and denoted as
$\alpha_{IR,n\ell,\overline{MS}}$ and $\alpha'_{IR,n\ell,S_{th_r}} \equiv 
\alpha'_{IR,n\ell,r}$. The 
$S_{th_r}$ entries are for $r=3, \ 6, \ 9, \ 4\pi$. As before, since the
two-loop IR zero is scheme-independent, we denote it simply as 
$\alpha_{IR,2\ell}$.}}
\begin{center}
\begin{tabular}{|c|c|c|c|c|c|c|c|c|c|c|c|c|} \hline\hline
$N$ & $N_f$ & $\alpha_{IR,2\ell}$ &
$\alpha_{IR,3\ell,\overline{MS}}$ & 
$\alpha'_{IR,3\ell,r=3}$  &
$\alpha'_{IR,3\ell,r=6}$  & 
$\alpha'_{IR,3\ell,r=9}$  &
$\alpha'_{IR,3\ell,r=4\pi}$  & 
$\alpha_{IR,4\ell,\overline{MS}}$ & 
$\alpha'_{IR,4\ell,r=3}$  & 
$\alpha'_{IR,4\ell,r=6}$  &
$\alpha'_{IR,4\ell,r=9}$  &
$\alpha'_{IR,4\ell,r=4\pi}$ \\
\hline
 2  &  7  &  2.83   & 1.05   & 1.07   & 1.11  & 1.21  & 1.45 
                    & 1.21   & 1.24   & 1.33  & 1.63  & complex \\
 2  &  8  &  1.26   & 0.688  & 0.693  & 0.706 & 0.731 & 0.781      
                    & 0.760  & 0.767  & 0.789 & 0.832 & 0.939 \\
 2  &  9  &  0.595  & 0.418  & 0.419  & 0.423 & 0.428 & 0.439     
                    & 0.444  & 0.446  & 0.450 & 0.458 & 0.472 \\
 2  & 10  &  0.231  & 0.196  & 0.196  & 0.197 & 0.197 & 0.199
                    & 0.200  & 0.200  & 0.201 & 0.202 & 0.203 \\
 \hline
 3  & 10  &  2.21   & 0.764  & 0.770  & 0.786 & 0.816 & 0.876       
                    & 0.815  & 0.822  & 0.844 & 0.885 & 0.978 \\
 3  & 11  &  1.23   & 0.578  & 0.581  & 0.588 & 0.602 & 0.627 
                    & 0.626  & 0.630  & 0.640 & 0.660 & 0.700 \\
 3  & 12  &  0.754  & 0.435  & 0.436  & 0.439 & 0.445 & 0.456
                    & 0.470  & 0.472  & 0.477 & 0.485 & 0.502 \\
 3  & 13  &  0.468  & 0.317  & 0.317  & 0.318 & 0.321 & 0.325
                    & 0.337  & 0.338  & 0.340 & 0.343 & 0.349 \\
 3  & 14  &  0.278  & 0.215  & 0.215  & 0.215 & 0.216 & 0.217
                    & 0.224  & 0.224  & 0.224 & 0.225 & 0.227 \\
 3  & 15  &  0.143  & 0.123  & 0.123  & 0.123 & 0.124 & 0.124
                    & 0.126  & 0.126  & 0.126 & 0.126 & 0.126 \\
 3  & 16  &  0.042  & 0.040  & 0.040  & 0.040 & 0.040 & 0.040
                    & 0.040  & 0.040  & 0.040 & 0.040 & 0.040 \\
\hline
 4  & 13  &  1.85   & 0.604  & 0.606  & 0.614 & 0.627 & 0.653
                    & 0.628  & 0.631  & 0.640 & 0.656 & 0.688 \\
 4  & 14  &  1.16   & 0.489  & 0.491  & 0.495 & 0.502 & 0.516
                    & 0.521  & 0.523  & 0.528 & 0.539 & 0.557 \\
 4  & 15  &  0.783  & 0.397  & 0.398  & 0.401 & 0.405 & 0.412
                    & 0.428  & 0.429  & 0.433 & 0.439 & 0.450 \\
 4  & 16  &  0.546  & 0.320  & 0.321  & 0.322 & 0.324 & 0.328   
                    & 0.345  & 0.346  & 0.348 & 0.351 & 0.357 \\
 4  & 17  &  0.384  & 0.254  & 0.254  & 0.255 & 0.256 & 0.258       
                    & 0.271  & 0.271  & 0.272 & 0.274 & 0.277 \\
 4  & 18  &  0.266  & 0.194  & 0.194  & 0.195 & 0.195 & 0.196
                    & 0.205  & 0.205  & 0.205 & 0.206 & 0.207 \\
 4  & 19  &  0.175  & 0.140  & 0.140  & 0.141 & 0.141 & 0.141
                    & 0.145  & 0.145  & 0.146 & 0.146 & 0.146 \\
 4  & 20  &  0.105  & 0.091  & 0.091  & 0.091 & 0.091 & 0.091
                    & 0.092  & 0.092  & 0.092 & 0.092 & 0.093 \\
 4  & 21  &  0.047  & 0.044  & 0.044  & 0.044 & 0.044 & 0.044
                    & 0.044  & 0.044  & 0.044 & 0.044 & 0.044 \\
\hline\hline
\end{tabular}
\end{center}
\label{betazero_fund_tanh}
\end{table}


\begin{table}
\caption{\footnotesize{Values of the IR zeros of $\beta_{\alpha}$ in the
$\overline{MS}$ scheme and $\beta_{\alpha'}$ after applying the $S_{sh_r}$
scheme transformation to the $\overline{MS}$ scheme, for an SU($N$) theory with
$N_f$ fermions in the fundamental representation, for $N=2,3,4$, calculated to
$n$-loop order and denoted as $\alpha_{IR,n\ell,\overline{MS}}$ and
$\alpha'_{IR,n\ell,S_{sh_r}} \equiv \alpha'_{IR,n\ell,r}$. The $S_{sh_r}$
entries are for $r=3, \ 6, \ 9, \ 4\pi$.  As before, since the two-loop IR zero
is scheme-independent, we denote it simply as $\alpha_{IR,2\ell}$.}}
\begin{center}
\begin{tabular}{|c|c|c|c|c|c|c|c|c|c|c|c|c|} \hline\hline
$N$ & $N_f$ & $\alpha_{IR,2\ell}$ &
$\alpha_{IR,3\ell,\overline{MS}}$ & 
$\alpha'_{IR,3\ell,r=3}$  &
$\alpha'_{IR,3\ell,r=6}$  & 
$\alpha'_{IR,3\ell,r=9}$  &
$\alpha'_{IR,3\ell,r=4\pi}$  & 
$\alpha_{IR,4\ell,\overline{MS}}$ & 
$\alpha'_{IR,4\ell,r=3}$  &
$\alpha'_{IR,4\ell,r=6}$  &
$\alpha'_{IR,4\ell,r=9}$  &
$\alpha'_{IR,4\ell,r=4\pi}$ \\
\hline
 2  &  7  &  2.83   & 1.05   & 1.05   & 1.03  & 0.998 & 0.953 
                    & 1.21   & 1.20   & 1.16  & 1.11  & 1.04  \\
 2  &  8  &  1.26   & 0.688  & 0.686  & 0.680 & 0.670 & 0.654      
                    & 0.760  & 0.757  & 0.747 & 0.732 & 0.7085\\
 2  &  9  &  0.595  & 0.418  & 0.418  & 0.416 & 0.413 & 0.409     
                    & 0.444  & 0.443  & 0.441 & 0.438 & 0.432 \\
 2  & 10  &  0.231  & 0.196  & 0.196  & 0.196 & 0.196 & 0.195
                    & 0.200  & 0.200  & 0.200 & 0.200 & 0.199 \\
 \hline
 3  & 10  &  2.21   & 0.764  & 0.762  & 0.754 & 0.742 & 0.723       
                    & 0.815  & 0.812  & 0.802 & 0.786 & 0.762 \\
 3  & 11  &  1.23   & 0.578  & 0.577  & 0.574 & 0.568 & 0.559 
                    & 0.626  & 0.6245 & 0.6195& 0.611 & 0.599 \\
 3  & 12  &  0.754  & 0.435  & 0.434  & 0.433 & 0.430 & 0.426
                    & 0.470  & 0.470  & 0.467 & 0.464 & 0.457 \\
 3  & 13  &  0.468  & 0.317  & 0.316  & 0.316 & 0.315 & 0.313
                    & 0.337  & 0.337  & 0.336 & 0.335 & 0.332 \\
 3  & 14  &  0.278  & 0.215  & 0.214  & 0.214 & 0.214 & 0.213
                    & 0.224  & 0.2235 & 0.223 & 0.223 & 0.222 \\
 3  & 15  &  0.143  & 0.123  & 0.123  & 0.123 & 0.123 & 0.123
                    & 0.126  & 0.126  & 0.126 & 0.126 & 0.125 \\
 3  & 16  &  0.042  & 0.040  & 0.040  & 0.040 & 0.040 & 0.040
                    & 0.040  & 0.040  & 0.040 & 0.040 & 0.040 \\
\hline
 4  & 13  &  1.85   & 0.604  & 0.602  & 0.599 & 0.593 & 0.583
                    & 0.628  & 0.626  & 0.622 & 0.615 & 0.603 \\
 4  & 14  &  1.16   & 0.489  & 0.488  & 0.486 & 0.483 & 0.477
                    & 0.521  & 0.520  & 0.517 & 0.513 & 0.505 \\
 4  & 15  &  0.783  & 0.397  & 0.397  & 0.396 & 0.394 & 0.390
                    & 0.428  & 0.428  & 0.426 & 0.423 & 0.419 \\
 4  & 16  &  0.546  & 0.320  & 0.320  & 0.319 & 0.318 & 0.316
                    & 0.345  & 0.345  & 0.344 & 0.343 & 0.340 \\
 4  & 17  &  0.384  & 0.254  & 0.253  & 0.253 & 0.253 & 0.252
                    & 0.271  & 0.271  & 0.271 & 0.270 & 0.268 \\
 4  & 18  &  0.266  & 0.194  & 0.194  & 0.194 & 0.194 & 0.193
                    & 0.205  & 0.205  & 0.204 & 0.204 & 0.2035 \\
 4  & 19  &  0.175  & 0.140  & 0.140  & 0.140 & 0.140 & 0.140
                    & 0.145  & 0.145  & 0.145 & 0.145 & 0.145 \\
 4  & 20  &  0.105  & 0.091  & 0.091  & 0.091 & 0.091 & 0.091
                    & 0.092  & 0.092  & 0.092 & 0.092 & 0.092 \\
 4  & 21  &  0.047  & 0.044  & 0.044  & 0.044 & 0.044 & 0.044
                    & 0.044  & 0.044  & 0.044 & 0.044 & 0.044 \\
\hline\hline
\end{tabular}
\end{center}
\label{betazero_fund_sinh}
\end{table}

\end{widetext} 


\begin{thebibliography}{99}

\bibitem{rg}
%
C. G. Callan, Phys. Rev. D {\bf 2}, 1541 (1970); K. Symanzik, Commun. Math.
Phys. {\bf 18}, 227 (1970).  Earlier related work includes M. Gell-Mann and
F. Low, Phys. Rev. {\bf 95}, 1300 (1954); N. N. Bogolubov and D. V. Shirkov,
Doklad. Akad. Nauk SSSR {\bf 103}, 391 (1955). See also K. Wilson, Phys. Rev. D
{\bf 3}, 1818 (1971).

\bibitem{nonsingular}
%
This makes implicit use of the property that the $\beta$ function is 
nonsingular in the interval $0 < \alpha \le \alpha_{IR}$. 

\bibitem{scc}
T. A. Ryttov and R. Shrock, arXiv:1206.2366.

\bibitem{ac}
T. Appelquist and J. Carazzone, Phys. Rev. D {\bf 11}, 2856 (1975). 

\bibitem{bvh}
T. A. Ryttov, R. Shrock, Phys. Rev. D {\bf 83}, 056011 (2011).

\bibitem{ps}
C. Pica, F. Sannino, Phys. Rev. D {\bf 83}, 035013 (2011). 

\bibitem{b1}
D. J. Gross and F. Wilczek, Phys. Rev. Lett. {\bf 30}, 1343 (1973);
H. D. Politzer, Phys. Rev. Lett. {\bf 30}, 1346 (1973); G. 't Hooft,
unpublished. 

\bibitem{b2}
W. E. Caswell, Phys. Rev. Lett. {\bf 33}, 244 (1974);
D. R. T. Jones, Nucl. Phys. B {\bf 75}, 531 (1974). 

\bibitem{gross75}
%
D. J. Gross, in R. Balian and J. Zinn-Justin, eds. {\it Methods in Field
Theory}, Les Houches 1975 (North Holland, Amsterdam, 1976).

\bibitem{dimreg}
G. 't Hooft, M. Veltman, Nucl. Phys. B {\bf 44}, 189 (1972). 

\bibitem{ms}
G. 't Hooft, Nucl. Phys. B {\bf 61}, 455 (1973).

\bibitem{msbar}
W. A. Bardeen, A. J. Buras, D. W. Duke, and T. Muta, Phys. Rev. D {\bf 18}, 
3998 (1978). 

\bibitem{b3}
O. V. Tarasov, A. A. Vladimirov, and A. Yu. Zharkov, Phys. Lett. B {\bf 93},
429 (1980); S. A. Larin and J. A. M. Vermaseren, Phys. Lett. B {\bf 303}, 334
(1993).

\bibitem{b4}
T. van Ritbergen, J. A. M. Vermaseren, and S. A. Larin, Phys. Lett. B {\bf
 400}, 379 (1997).

\bibitem{b1b2}
D. J. Gross and F. Wilczek, Phys. Rev. D {\bf 8}, 3633 (1973); 
Phys. Rev. D {\bf 9}, 980 (1974); H. D. Politzer, Phys. Rept. {\bf 14}, 124
(1974). 

\bibitem{bethke}
S. Bethke, Eur. Phys. J. C {\bf 64}, 689 (2009).

\bibitem{dyson}
%
One may recall the proof that amplitudes and quantities such as $\beta$ and
$\gamma_m$ are nonanalytic at $\alpha=0$ in a gauge theory.  Assume the
contrary; then one could analytically continue these quantities to real $\alpha
< 0$, but this would correspond to an imaginary gauge coupling, which would
violate the unitarity of the theory.

\bibitem{cdg}
%
Instanton effects on $\beta$ were estimated in QCD in C. G. Callan,
R. F. Dashen, and D. J. Gross, Phys. Rev. D {\bf 17}, 2717 (1978); Phys. Rev. D
{\bf 20}, 3279 (1979) and were found to increase the magnitude of $\beta$.

\bibitem{thooft77}
G. 't Hooft, in {\it The Whys of Subnuclear Physics, Proc. 1977
Erice Summer School}, ed. A. Zichichi (Plenum, New York, 1979), p. 943. 

\bibitem{stevenson}
P. M. Stevenson, Phys. Rev. D {\bf 23}, 2916 (1981); Nucl. Phys. B {\bf 203},
472 (1982); Nucl. Phys. B {\bf 231}, 65 (1984). 

\bibitem{braaten}
E. Braaten and J. P. Leveille, Phys. Rev. D {\bf 24}, 1369 (1981). 

\bibitem{blm}
S. J. Brodsky, G. P. Lepage, and P. B. MacKenzie, Phys. Rev. D {\bf 28}, 228
(1983); S. J. Brodsky and H. J. Lu, Phys. Rev. D {\bf 51}, 3652 (1995); 
S. J. Brodsky and P. Huet, Phys. Lett. B {\bf 417}, 145 (1998). 

\bibitem{brodskyst}
%
S. J. Brodsky and X.-G. Wu, Phys.Rev. D {\bf 85}, 034038 (2012);
arXiv:1203.5312; arXiv:1204.1405. 

\bibitem{kataev}
A. V. Garkusha and A. L. Kataev, Phys. Lett. B {\bf 705}, 400 (2011). 

\bibitem{casimir}
%
The Casimir invariants $C_R$ and $T_R$ are defined as $\sum_a \sum_j {\cal
D}_R(T_a)_{ij} {\cal D}_R(T_a)_{jk} = C_R \delta_{ik}$ and $\sum_{i,j} {\cal
D}_R(T_a)_{ij} {\cal D}_R(T_b)_{ji} = T_R \delta_{ab}$, where $R$ is the
representation and $T_a$ are the generators of $G$, so that for SU($N_c$),
$C_A=N_c$ for the adjoint ($A$) and $T_{fund}=1/2$ for the
fundamental representation, etc. $C_f$ denotes $C_R$ for the fermion 
representation. 

\bibitem{nfreal}
%
Here and elsewhere, when expressions are given for $N_f$ that evaluate
to non-integral real values, it is understood that they are formal and are 
interpreted via an analytic continuation of $N_f$ from physical nonnegative 
integer values to real numbers. 


\bibitem{nonpertzero}
%
We focus here on an IR zero of the perturbative $\beta$ function. 
A nonperturbative zero in $\beta$ has been discussed in
S. J. Brodsky, G. F. de T\'eramond, and A. Deur, Phys. Rev. D {\bf 81}, 096010
(2010); and M. Creutz, Acta Phys. Slovaca {\bf 61}, 1 (2011). 

\bibitem{bz}
T. Banks and A. Zaks, Nucl. Phys. B {\bf 196}, 189 (1982).

\bibitem{conf}
%
For a chiral gauge theory with requisite fermion content, there is also the
alternate possibility that the theory may confine and produce massless
composite fermions, as discussed in G. 't Hooft, {\it Recent Developments in
Gauge Theories, Carg\'ese Summer Institute, 1979} (Plenum, New York, 1980),
p. 135; recent studies include T. Appelquist, A. Cohen, M. Schmalz, and
R. Shrock, Phys. Lett. B {\bf 459}, 235 (1999); T. Appelquist and F. Sannino,
Phys.Rev. D {\bf 61}, 125009 (2000), and references therein.

\bibitem{wtc}
B. Holdom, Phys. Lett. B {\bf 150}, 301 (1985); K. Yamawaki, M. Bando, and 
K. Matumoto, Phys. Rev. Lett.  {\bf 56}, 1335 (1986); 

\bibitem{chipt}
T. Appelquist, D. Karabali, and L. C. R. Wijewardhana, Phys. Rev. Lett. {\bf
57}, 957 (1986); T. Appelquist and L. C. R. Wijewardhana, Phys. Rev. D
{\bf 35}, 774 (1987);  Phys.  Rev. D {\bf 36}, 568 (1987).

\bibitem{dscor}
%
T. Appelquist, K. D. Lane, and U. Mahanta, Phys.Rev.Lett. {\bf 61}, 1553
T. Appelquist and S. Selipsky, Phys. Lett. B {\bf 400}, 364 (1997);
S. J. Brodsky and R. Shrock, Phys. Lett. B {\bf 666}, 95 (2008).  

\bibitem{tc}
%
The latter is the case in QCD and technicolor theories:
S. Weinberg, Phys. Rev. D {\bf 19}, 1277 (1979);
L. Susskind, Phys. Rev. D {\bf 20}, 2619 (1979).

\bibitem{bs1}
M. Harada, M. Kurachi and K. Yamawaki, Phys. Rev. D {\bf 68}, 076001 (2003); 
Phys. Rev. D {\bf 70}, 033009 (2004).

\bibitem{bs2}
M. Kurachi and R. Shrock, M. Kurachi, JHEP {\bf 12}, 034 (2006); 
Phys. Rev. D {\bf 74}, 056003 (2006).

\bibitem{lgt1}
T. Appelquist, G. Fleming, and E. Neil, Phys. Rev. Lett. {\bf 100},
171607 (2008); Phys. Rev. D {\bf 79}, 076010 (2009); T. Appelquist et al.,
Phys. Rev. Lett. {\bf 104}, 071601 (2010); T. Appelquist et al., Phys. Rev. D
{\bf 84}, 054501 (2011); A. Deuzeman, M. P. Lombardo, and E. Pallante,
Phys. Lett. B {\bf 670}, 41 (2008); Phys. Rev. D {\bf 82}, 074503 (2010);
A. Hasenfratz, Phys. Rev. Lett. {\bf
108}, 061601; Y. Aoki et al., arXiv:1202.4916; 

\bibitem{lgt2}
Z. Fodor et al., Phys. Lett. B {\bf 681}, 353 (2009); M. Hayakawa et al.,
Phys. Rev. D {\bf 83}, 074509 (2011); X.-Y. Jin and R. D. Mawhinney,
arXiv:1203.5855.  

\bibitem{lgtrev}
%
For recent reviews of this and other gauge theories on the 
lattice, see talks in https://latt11.llnl.gov; \goodbreak 
http://lqcd.fnal.gov/(tilde)eneil/lat-exp-2011; and \goodbreak 
http://www.kmi.nagoya-u.ac.jp/workshop/SCGT12Mini. 

\bibitem{as}
%
See, e.g., T. Appelquist and J. Terning, Phys. Rev. D {\bf 50}, 2116 (1994);
T. Appelquist and R. Shrock, Phys. Lett. B {\bf 548}, 204 (2002);
Phys. Rev. Lett. {\bf 90}, 201801 (2003); T. Appelquist, M. Piai, and 
R. Shrock, Phys. Rev. D {\bf 69}, 015002 (2004); N. D. Christensen and
R. Shrock, Phys. Rev. Lett. {\bf 94}, 241801 (2005). 

\bibitem{bfs}
%
An analogous study of the effect of higher-loop terms in the $\beta$ function
on the UV to IR evolution of a supersymmetric gauge theory has been carried out
in T. A. Ryttov, R. Shrock, Phys. Rev. D {\bf 85}, 076009 (2012).  See also 
T. A. Ryttov and F. Sannino, Phys. Rev. D {\bf 78}, 065001 (2008); 
C. Pica and F. Sannino, Phys.Rev. D {\bf 83}, 116001 (2011). 

\bibitem{mom}
W. Celmaster and R. J. Gonsalves, Phys. Rev. D {\bf 20}, 1420 (1979); a recent
study is J. A. Gracey, Phys. Rev. D {\bf 84}, 085011 (2011). 

\bibitem{gardi}
E. Gardi, G. Grunberg and M. Karliner, JHEP {\bf 07}, 007 (1998).

\bibitem{sanrev}
D. D. Dietrich, F. Sannino, and K. Tuominen, Phys. Rev. D {\bf 72}, 
055001 (2005); R. Foadi, M. T. Frandsen, T. A. Ryttov, and 
F. Sannino, Phys. Rev. D {\bf 76}, 055005 (2007); F. Sannino, Acta
Phys. Polon. B {\bf 40}, 3533 (2009). 

\end{thebibliography}
\end{document}